\documentclass[11pt]{article}
\usepackage{style}

\begin{document}

\title{Fast spatial simulation of extreme high-resolution radar precipitation data using INLA}

\date{}

\author{Silius M. Vandeskog\\
  \small{Department of Mathematics, The Norwegian University of Science and Technology (NTNU),}\\
  \small{and The Norwegian Computing Center (NR)}\\
  and \\
  Rapha{\"e}l Huser \\
  \small{Statistics program, CEMSE Division, King Abdullah University of Science and Technology
    (KAUST)} \\
  and \\
  Oddbjørn Bruland \\
  \small{Department of Civil and Environmental Engineering, The Norwegian University of Science and
    Technology (NTNU)} \\
  and \\
  Sara Martino \\
  \small{Department of Mathematics, The Norwegian University of Science and Technology (NTNU)}\\
}

\maketitle

\bigskip
\begin{abstract}
Aiming to deliver improved precipitation simulations for hydrological impact assessment studies,
  we develop a methodology for modelling and simulating high-dimensional spatial precipitation
  extremes, focusing on both their marginal distributions and tail dependence structures.  Tail
  dependence is crucial for assessing the consequences of extreme precipitation events, yet most
  stochastic weather generators do not attempt to capture this property.
  The spatial distribution of precipitation occurrences is modelled with four competing models, 
  while the spatial distribution of nonzero extreme precipitation intensities are modelled with a latent Gaussian version of the spatial conditional extremes model.
  Nonzero precipitation marginal distributions are modelled using latent Gaussian models with gamma and generalised
  Pareto likelihoods.
  Fast inference is achieved using integrated nested Laplace approximations
  (INLA).
  We model and simulate spatial precipitation extremes in Central Norway, using 13 years of hourly radar
  data with a spatial resolution of \(1 \times 1\)~km\(^2\), 
  over an area of size \(6461\)~km\(^2\),
  to describe the behaviour of extreme precipitation over a small drainage area. Inference on this high-dimensional data set is achieved within hours, and the simulations
  capture the main trends of the observed precipitation well.
\end{abstract}

\noindent%
\vfill

\maketitle

\section{Introduction}%
\label{sec:introduction}

Europe is currently experiencing one of its most flood-intense periods within the last 500 years
\citep{bloeschl20_curren_europ_flood_rich_period}, and floods are projected to become more frequent
and damaging in the future due to ongoing climate changes
\citep{yin18_large_increas_global_storm_runof, allan20_advan_under_large_scale_respon}. Thus, flood
mitigation has the potential of avoiding numerous fatalities and large economical losses
\citep{jongman18_effec_adapt_risin_flood_risk}. This mitigation depends upon design criteria for
critical infrastructures such as dams, bridges, culverts and drainage structures in urban
areas. Among the most important of these criteria is the design flood, which is estimated using
statistical and hydrological modelling, where precipitation often is the most important input. This
precipitation input is typically provided by rain gauge observations, interpolated observational
data sets or climate projections from general circulation models and regional climate models
\citep{hanssen-bauer15_klima_norge, giorgi19_thirt_years_region_climat_model}. However,
precipitation is a localised phenomenon with much space-time variability, which the rain gauge
observations, interpolated data sets and climate projections are unable to capture due to
computational constraints and sparsity of observations in space and time
\citep{westra14_futur_chang_to_inten_frequen, lopez-cantu20_uncer_futur_u}. The observational data
sets may also be too short in time to fully capture the risks and consequences of floods, as the
most devastating extreme weather events with high flood risk may simply have not happened yet.
Stochastic weather generators, which are methods for simulating climate data,
have therefore become important tools for climate impact assessment methods, such as design flood
estimation, where simulated precipitation is used as input for continuous streamflow
simulation~\citep{BOUGHTON2003309}.  Such simulated climate data allow for better exploration of
complex weather phenomena by providing longer time series, or by capturing important small-scale
spatio-temporal variability that happens ``inside the grids'' of too coarse climate projections and
interpolated data sets \citep{AllardEtAl2014Disaggregatingdailyprecipitations,
  ailliot15_stoch_weath_gener, maraun18_statis_downs_bias_correc_climat_resear}.

The spatio-temporal distribution of precipitation over a catchment influences the catchment's flood
peak response and should therefore be included in the flood frequency analysis and subsequent design
flood calculations~\citep{wilson1979influence, Shakti11081703_Spatial_Distribution_Radar_Japan,
  Storm_Dir_Perez, ToulemondeEtAl2020Space-TimeSimulationsExtreme, GHIMIRE2021126683}. Most
stochastic weather generators are purely temporal, but multiple spatio-temporal or multi-site
generators also exist \citep{Wilks1998Multisitegeneralizationdaily,
  buishand2001,
  wheater05_spatial_tempor_rainf_model_flood_risk_estim, BenoitEtAl2017Generatingsyntheticrainfall,
  peleg17_advan_stoch_weath_gener_simul, benoit18_stoch_rainf_model_sub_scale}.
  Additionally, most
stochastic weather generators focus on simulating ``non-extreme'' precipitation, but the topic of
modelling (or even projecting) extreme precipitation, also in the spatio-temporal setting, has received increased
attention in the last years \citep{BaxevaniEtAl2015spatiotemporalprecipitationgenerator,
  WinterEtAl2020Eventgenerationprobabilistic, WilcoxEtAl2021StochastormStochasticRainfall,zhong2022spatialmodelingfutureprojection,
  BoutignyEtAl2023metaGaussiandistribution, JiEtAl2024Implementinggenerativeadversarial}. However,
these weather generators mainly focus on capturing the marginal behaviour of extremes, while giving
less focus to describing spatio-temporal tail dependence properties. There can be considerable
differences between regular dependence and tail dependence \citep{sibuya60_bivar_extrem_statis}, and
the tail depence properties of precipitation should also be taken into account as these can be of
utmost importance for, e.g., flood impact assessments \citep{LeEtAl2018Dependencepropertiesspatial}.
\citet{ToulemondeEtAl2020Space-TimeSimulationsExtreme} describe why spatio-temporal simulations of
extreme precipitation are important for mitigating flood risks in small and urban catchments, while
highlighting the lack of stochastic weather generators that are suitable for this
task. \citet{EvinEtAl2018Stochasticgenerationmulti} simulate precipitation using a transformed
multivariate autoregressive model with Gaussian or Student-\textit{t} distributed innovation terms,
for modelling precipitation using both asymptotic dependence and asymptotic independence.
\citet{palacios-rodriguez20_gener_paret_proces_simul_space_extrem_event} and
\citet{GuevaraEtAl2023DirectSamplingSpatially} simulate high-resolution spatio-temporal
precipitation extremes by resampling transformations of observed extreme events. 
This effectively captures the spatio-temporal tail dependence structure, but it also prevents the generation of events with behaviours that differ from those already observed.
\citet{richards22_model_extrem_spatial_aggreg_precip_condit_method,
  richards23_joint_extrem_spatial_aggreg_precip} develop promising spatial simulations of extreme
hourly precipitation, but their method is based on an inefficient inference scheme that becomes
troublesome for higher-dimensional problems.

Hydrological impact assessment studies over small or urban catchments often require detailed
precipitation information at high spatio-temporal resolutions, as lower-resolution simulations
smooth out the distribution of precipitation too much and therefore fail to capture the extreme
behaviour and large degrees of intermittency of high-resolution precipitation data
\citep{TetzlaffEtAl2005Significancespatialvariability,
  VenezianoEtAl2006Multifractalityrainfallextremes, PaschalisEtAl2014temporalstochasticmodeling,
  ToulemondeEtAl2020Space-TimeSimulationsExtreme}. However, most stochastic weather generators rely
on observation data from rain gauges, which generally are too sparsely located in both space and
time to capture the properties of extreme precipitation well \citep{LengfeldEtAl2020Useradardata}.
A promising alternative is the use of weather radar observations, which have a much finer
spatio-temporal resolution. Weather radars are known to be unable to capture marginal distributions as well
as rain gauges, but they provide reliable descriptions of the spatio-temporal dependence of
precipitation data \citep{BardossyEtAl2017Combinationradardaily,
  Ochoa-RodriguezEtAl2019ReviewRadar-RainGauge, bournas22_deter_z_r_relat_spatial}.
While several studies have used radar data to investigate the spatial distribution of precipitation
and its influence on runoff hydrographs in urban
areas~\citep[e.g.,][]{LOWE2014397,ochoa-rodriguez15_impac,rico-ramirez15_quant,hess-21-3859-2017}
fewer studies are found on the larger catchment scales, such as those studied by
\citet{tramblay2011impact} and \citet{onate18_tempor_spatia_precip_pattern}. Attempts at spatial or
spatio-temporal modelling of extreme precipitation, based on radar data, have only been
made recently~\citep{OvereemEtAl2009Extremerainfallanalysis, MarraEtAl2022Coastalorographiceffects,
  ANSHSRIVASTAVA2023129902}.

Recently, a new research frontier, so-called “gamification” or serious games, has emerged as a way to encourage community awareness and engagement on various issues---including planning activities in disaster scenarios such as floods \citep{Game2, Game1, Game3}. 
Projects like the World of Wild Waters (WoWW) \citep{WoWW}, which our work is a part of, aim at providing the user with  an immersive  experience, using state-of-the-art hydrological models and Virtual Reality \citep{Gebray2024}. To do so, high resolution,  realistic scenarios of extreme precipitation fields  are needed.

In this paper we build upon the work of \citet{VandeskogEtAl2024efficientworkflowmodelling} to develop a framework for computationally efficient high-dimensional spatial
modelling and simulation of extreme precipitation, that captures both the marginal and the joint tail
behaviour and the intermittency of high-resolution precipitation data.  The framework is applied to 
simulate spatial precipitation extremes using a data set of high-resolution hourly precipitation
data from a weather radar in Norway. The overarching goal of this work is to establish efficient
methods for generating statistically realistic extreme precipitation scenarios that could serve as input  for hydrological models. Outcomes from such hydrological models could  provide a basis for flood frequency analysis, estimation of design floods and scenarios for serious games such as in the WoWW project. Spatio-temporal simulation is outside
the scope of the paper, but extending our framework to space-time should be fairly straightforward
(see Section~\ref{sec:conclusion}). 

Our novel stochastic weather generator for spatial precipitation extremes builds upon extreme value
theory \citep[e.g.,][]{davison15_statis_extrem}, which has shown great success at modelling and
assessing environmental risks such as extreme temperature
\citep{castro-camilo21_bayes_space_time_gap_fillin, simpson21_condit_model_spatio_tempor_extrem},
precipitation \citep{buishand2008, huser14_space_time_model_extrem_event,
  opitz18_inla_goes_extrem,
  richards22_model_extrem_spatial_aggreg_precip_condit_method} and wind
\citep{castro-camilo19_splic_gamma_gener_paret_model}.
The extreme value theory literature is heavily concerned with the modelling of those tail dependence
properties that \citet{ToulemondeEtAl2020Space-TimeSimulationsExtreme} highlight as a missing focus
in the stochastic weather generator literature. An important tail dependence property, also known as 
extremal dependence, can be described using conditional exceedance probabilities, as follows. Given
a spatial random field, \(X(\bm s)\) with \(\bm s \in \mathcal S \subset \mathbb R^2\), we consider
the conditional exceedance probability
\begin{equation}
  \chi_p(\bm s_1, \bm s_2) = \text{P}(X(\bm s_1) > F_{\bm s_1}^{-1}(p)
  \mid X(\bm s_2) > F_{\bm s_2}^{-1}(p)),
\end{equation}
where \(F_{\bm s}^{-1}\) is the quantile function of \(X(\bm s)\), and we then define the tail correlation
coefficient as \(\chi(\bm s_1, \bm s_2) = \lim_{p \rightarrow 1} \chi_p(\bm s_1, \bm s_2)\). Two random
variables, \(X(\bm s_1)\) and \(X(\bm s_2)\), are called asymptotically dependent if
\(\chi(\bm s_1, \bm s_2) > 0\), and asymptotically independent otherwise.  Experience has shown that
environmental data often exhibit weakening dependence as events become more extreme
\citep{huser21_max_infin_divis_model_infer_spatial_extrem}, i.e., \(\chi_p\) continuously decreases
as \(p \rightarrow 1\). However, classical models for spatial extremes are often based on max-stable
processes \citep{davison19_spatial_extrem}, which focus on modelling pointwise maxima and assume
that \(\chi\) is positive while \(\chi_p\) is nearly constant with level \(p \approx 1\). Thus,
alternative models that focus on capturing the so-called subasymptotic dependence structure
explained by \(\chi_p\) are crucial for correctly assessing the risks of spatial extremes in
environmental data. A spatial process used for modelling climate data should also be able to exhibit
both asymptotic dependence at short distances and asymptotic independence at large distances, but
most classical extreme models are unable to describe nontrivial changes in the asymptotic dependence
class as a function of distance \citep{huser22_advan_statis_model_spatial_extrem,
  huser24_model_spatial_extrem_envir_data_scien}. This has led to a surge of new models for spatial
extremes with more flexible subasymptotic and asymptotic dependence structures, including inverted
max-stable processes and max-mixture models \citep{wadsworth12_depen_model_spatial_extrem},
max-infinitely divisible processes \citep{huser21_max_infin_divis_model_infer_spatial_extrem},
scale-mixture models \citep{huser17_bridg_asymp_indep_depen_spatial,
  engelke19_extrem_depen_random_scale_const, huser19_model_spatial_proces_with_unknow}, kernel
convolution models \citep{krupskii22_model_spatial_tail_depen_cauch_convol_proces} and the spatial
conditional extremes model \citep{wadsworth22_higher_spatial_extrem_singl_condit}.

Statistical modelling of spatial dependence often leads to computationally demanding inference. This
is particularly true for spatial extreme value models, where many of the most popular models
have to rely on low-dimensional composite likelihood methods to achieve computationally tractable
inference \citep{padoan10_likel_based_infer_max_stabl_proces,
  castruccio16_high_order_compos_likel_infer}. The Gaussian random field is popular in traditional
spatial and spatio-temporal statistics, as it has nice theoretical properties while allowing for
fast and realistic modelling of complex processes \citep{gelfand10_handb_spatial_statis}. In
particular, the latent Gaussian modelling framework has shown great success within a large range of
applications \citep{banerjee14_hierar_model_analy_spatial_data}, by yielding flexible and realistic
models that utilise assumptions of Gaussianity and conditional independence for performing fast
inference using integrated nested Laplace approximations
\citep[INLA;][]{rue09_approx_bayes_infer_laten_gauss}. Yet, latent Gaussian models have not achieved
similar success for modelling spatial extremes, as their dependence structures are unsuitable for
most classical spatial extreme value models \citep{davison12_statis_model_spatial_extrem}. However,
Gaussian dependence structures are becoming more suitable for some newer breeds of spatial extreme
value models, such as the spatial conditional extremes model. Indeed, the model only requires a few minor
alterations to become a latent Gaussian model, which makes it possible to perform fast
high-dimensional inference with INLA \citep{SimpsonEtAl2023Highdimensionalmodeling,VandeskogEtAl2024efficientworkflowmodelling}.

We model spatial precipitation extremes with the spatial conditional extremes model and INLA, and we develop new empirical
diagnostics and parametric models for describing components of the spatial conditional extremes
model, as well as improved models for the marginal distributions and a new methodology for
describing precipitation zeros within this framework.
The spatial conditional extremes model extends the (multivariate) conditional extremes model of \citet{heffernan04_condit_approac_multiv_extrem_values} and \citet{heffernan07_limit_laws_random_vector_with_extrem_compon}. It describes the distribution of a spatial random field
\({\{Y(\bm s)\}}_{\bm s \in \mathcal S \subset \mathbb R^2}\), with Laplace marginal distributions,
given that it exceeds a large threshold \(\tau\) at some preselected location
\(\bm s_0 \in \mathcal S\). The model assumes that, for \(\tau\) large enough, the process
\([Y(\bm s) \mid Y(\bm s_0) = y_0]\), with \(y_0 > \tau\), is approximately equal in distribution to
a spatial random field that only depends on \(y_0\) through a location parameter
\(a(\bm s; \bm s_0, y_0)\) and a scale parameter \(b(\bm s; \bm s_0, y_0) > 0\). An important part
of the modelling process is therefore the choice of a suitable class of functions for \(a(\cdot)\) and
\(b(\cdot)\), and of a threshold \(\tau\) that is high enough to yield little model bias, but
also small enough to efficiently utilise the data. To the best of our knowledge, this threshold
selection problem has not yet attracted much focus in the spatial extremes literature. In this paper we develop new
empirical diagnostics for finding reasonable values of the threshold \(\tau\) and the forms of
\(a(\cdot)\) and \(b(\cdot)\), and we propose a new class of parametric functions for \(a(\cdot)\)
and \(b(\cdot)\) that can provide suitable fits to data at much lower thresholds than used
previously \citep[e.g.,][]{VandeskogEtAl2024efficientworkflowmodelling}, thus allowing
us to utilise more of the data for more efficient inference, without too much model bias.

To fit the spatial conditional extremes model, one must first standardise the data to have Laplace
marginal distributions. However, the marginal distributions of hourly precipitation contain a point
mass at zero, which makes it impossible to directly transform them to the Laplace scale using the
probability integral transform. \citet{richards22_model_extrem_spatial_aggreg_precip_condit_method,
  richards23_joint_extrem_spatial_aggreg_precip} solve this problem by censoring all zeros, but this
leads to less efficient inference techniques such as low-dimensional composite likelihoods, and it
cannot easily be combined with the INLA framework. Inspired by the so-called Richardson-type
stochastic weather generators \citep{richardson81_stoch_simul_daily_precip_temper_solar_radiat}, we
instead propose to model the conditional extremes of nonzero precipitation intensity, while
separately modelling the distribution of precipitation occurrences in space. We then combine the two
models together to describe the full distribution of spatial conditional precipitation extremes. This is a
common method for simulating precipitation \citep{SternEtAl1984ModelFittingAnalysis, SchleissEtAl2014Stochasticsimulationintermittent,
  EvinEtAl2018Stochasticgenerationmulti, WilcoxEtAl2021StochastormStochasticRainfall}, but to the
best of our knowledge, it has not previously been used in conjunction with the spatial
conditional extremes model.
We propose multiple competing models for the distribution of the conditional precipitation
occurrences. The probit model is a common regression model for binary data
\citep{fahrmeir13_regression, verdin15_coupl_stoch_weath_gener_using}, and we use 
both the standard version and a spatial version of it to model the precipitation occurrences. We show
that both probit models are latent Gaussian models, and we perform fast inference for them using INLA\@.
However, our probit models produce occurrence values that are independent of the precipitation
intensity values, which is unrealistic. The probit model also struggles to capture some other
important spatio-temporal properties of smooth high-resolution precipitation data.
Thus, we propose an additional, third model, denoted the threshold model, which is designed to
capture the dependence in the precipitation intensity process and to better capture the spatial
smoothness properties of precipitation occurrences. For better baseline comparisons, we also propose
an occurrence model in which ``no precipitation'' is interpreted as a tiny but positive amount of
precipitation, or in other words, that it always rains.

To transform nonzero precipitation data onto the Laplace scale, we must first estimate their
marginal distributions in space and time. In the spatial conditional extremes literature, this is
commonly achieved using the empirical distribution functions at each location, possibly combined
with a generalised Pareto (GP) distribution for describing the upper tail
\citep{simpson21_condit_model_spatio_tempor_extrem,
  richards22_model_extrem_spatial_aggreg_precip_condit_method,
  wadsworth22_higher_spatial_extrem_singl_condit,
  shooter22_multiv_spatial_condit_extrem_extrem_ocean_envir}. However, the empirical distribution
function can be unsuitable if the marginal distribution of the data varies in space and time, which
is often the case for precipitation and other climate variables. Since both the total amount and the
spatial distribution of precipitation are important properties for assessing flood risk, we here
focus equally on describing properties of the marginal and the spatial precipitation
distributions. Therefore, following \citet{opitz18_inla_goes_extrem} and
\citet{castro-camilo19_splic_gamma_gener_paret_model}, we apply a complex spatio-temporal model
based on two different latent Gaussian models for describing the marginal distributions. The first
model describes the bulk of the data using a gamma likelihood, while the second model describes the
upper tail using a GP likelihood.

To sum up, in this paper we develop a framework for modelling and simulating extreme precipitation
in space, based on latent Gaussian models and the spatial conditional extremes model.
This is
applied for creating precipitation extremes scenarios using a high-resolution data set of hourly
precipitation from a weather radar in Norway.
The modelling framework builds upon previous work by \citet{SimpsonEtAl2023Highdimensionalmodeling} and \citet{VandeskogEtAl2024efficientworkflowmodelling},
but we extend it by proposing novel empirical diagnostics for choosing model components of the spatial conditional extremes model, and
we use these for proposing new parametric functions for the model components, 
which allow for better data utilisation through a lower extremal threshold.
We also propose a new method for modelling precipitation zeros within the spatial conditional extremes framework, by
separately modelling precipitation occurrences and
intensities. The spatial distribution of
the precipitation occurrences is described using four competing models, while the marginal distributions
of nonzero precipitation are modelled by merging two latent Gaussian models, with a gamma likelihood
and a GP likelihood, respectively. We employ a latent Gaussian model version of the spatial
conditional extremes model for describing the extremal dependence of the nonzero precipitation.
The remainder of the paper is
organised as follows: The weather radar data are presented in Section~\ref{sec:data}. Then,
Section~\ref{sec:model_framework} describes our general framework for modelling spatial extreme
precipitation, and in Section~\ref{sec:case_study}, we apply our framework for modelling and
simulating extreme precipitation using the chosen radar data. The paper concludes with a final
discussion in Section~\ref{sec:conclusion}.

\section{Data}%
\label{sec:data}

The Rissa radar is located at an elevation of 616 meters above sea level at the Fosen peninsula in
Central Norway.  Further description of the radar specifications and properties are given by
\citet{10.2166/nh.2010.011}, who also analyse the accuracy of the radar observations at different
ranges. The Norwegian Meteorological Institute processes the observed reflectivity data and uses
them to create gridded \(1 \times 1\)~km\(^2\) resolution maps of estimated hourly precipitation,
measured in mm/h \citep{elo12_correc_quant_radar_data}. These precipitation maps are freely
available, dating back to 1 January 2010, from an online weather data archive
(\url{https://thredds.met.no}).

\begin{figure}
  \centering
  \hfill
  \includegraphics[width=.3\linewidth]{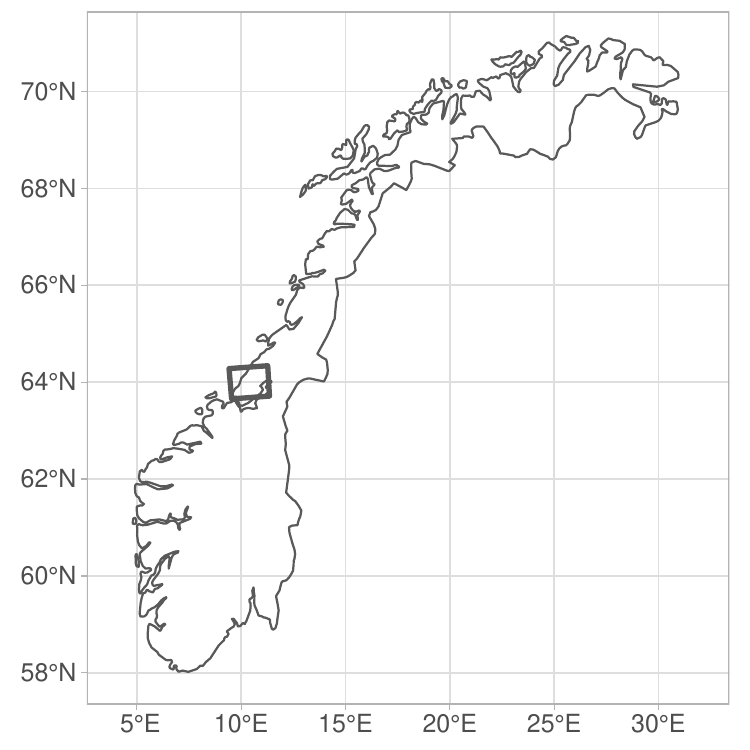}
  \hfill
  \includegraphics[width=.5\linewidth]{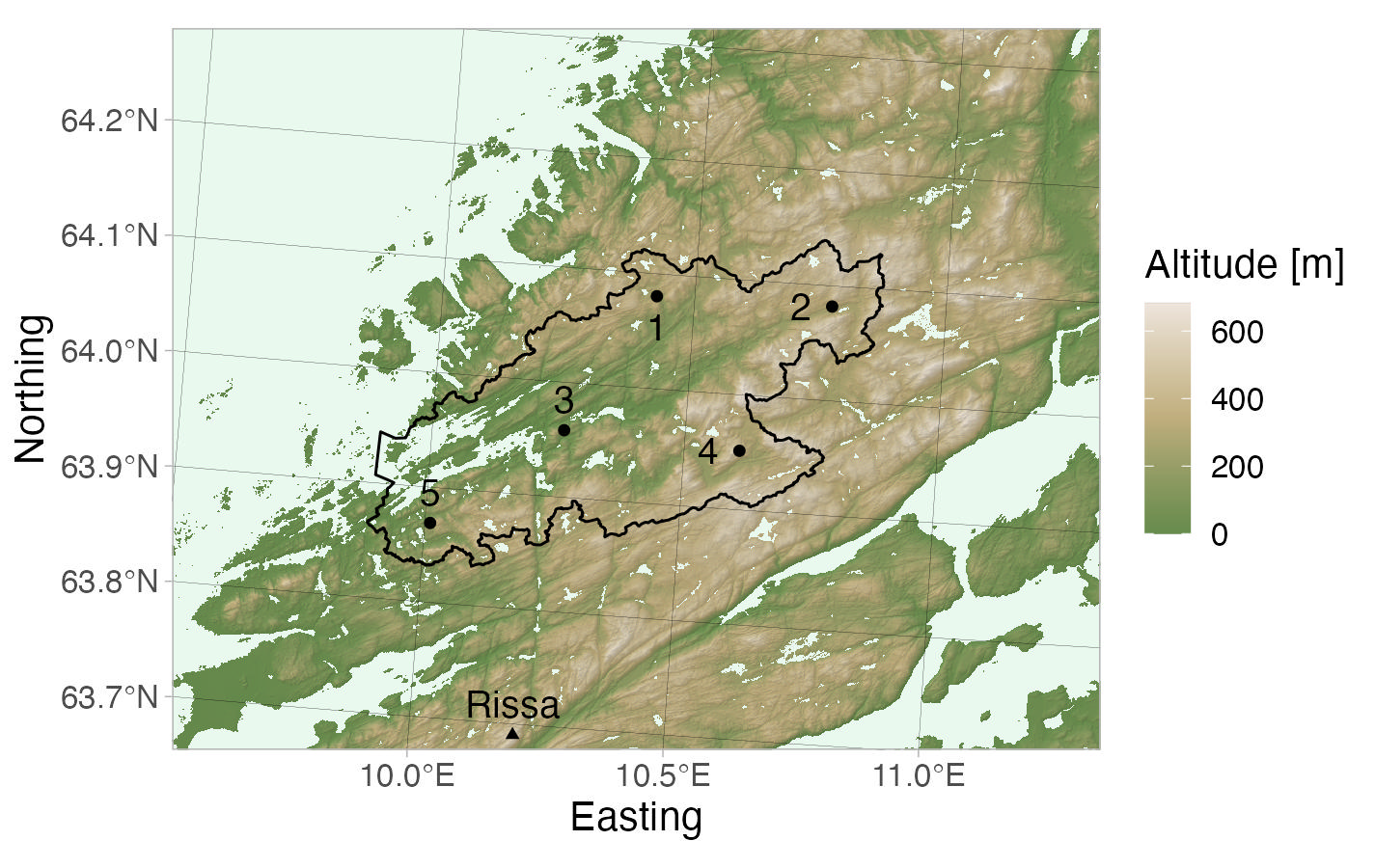}
  \hfill
  \caption{Left plot: Map of Norway with a square defining the spatial domain \(\mathcal S\). Right
    plot: Elevation map of \(\mathcal S\). The Rissa radar is displayed at the bottom of the map,
    and the five conditioning sites used in Section~\ref{sec:case_study} are enumerated and marked
    by black circles. A drainage area in Åfjorden municipality is displayed using a black polygon.}%
  \label{fig:data_domain}
\end{figure}

We use the radar precipitation maps for modelling and simulating extreme hourly precipitation over a
840 km\(^2\) drainage area in Åfjorden, located close to the Rissa radar. To achieve this, we
download all data for 2010--2022 inside a spatial domain \(\mathcal S\) of size
\(91 \times 71 (= 6461)\)~km\(^2\), centred around the drainage area. Figure~\ref{fig:data_domain}
displays the domain \(\mathcal S\), the locations of the Rissa radar and the drainage area of
interest. The spatial conditional extremes model allows one to model and simulate extremes occurring
at any site of interest, by conditioning on that site experiencing extreme behaviour (see
Section~\ref{sec:model_framework} for more details). For the sake of illustration, we choose five
such conditioning sites, somewhat equally spaced throughout the water catchment, for modelling and
simulating extreme precipitation in Section~\ref{sec:case_study}.
These sites are also displayed in Figure~\ref{fig:data_domain}. There are considerable differences
between extreme summer precipitation and extreme winter precipitation in Norway
\citep{dyrrdal15_bayes_hierar_model_extrem_hourl_precip_norway}, and we therefore choose to only
model summer precipitation from June, July and August, which is when most of the intense
precipitation events occur in Norway. There are some distortions in the data close to the Rissa
radar, so we remove all observations from locations that are within \(5\)~km from the radar.

It can be difficult to distinguish between little and no precipitation using reflectivity data, and
the estimated precipitation data contain both exact zeros and values with magnitudes as small as
\(10^{-5}\)~mm/h. In Section~\ref{sec:supp_data_exploration} of the Supplementary Material,
we show that there are large differences in the proportions of exact zeros for different times, due
to an upgrade of the weather radar in 2018, but that the proportion of observations smaller than
\(0.1\)~mm/h is approximately constant in time. We therefore round every observation smaller than
\(0.1\)~mm/h down to zero.

\section{Modelling framework}%
\label{sec:model_framework}

\subsection{Model overview}
We model the spatial extremal dependence structure of the hourly precipitation process,
\(X_t(\bm s)\), at location \(\bm s \in \mathcal S \subset \mathbb R^2\) and time
\(t \in \mathcal T \subset \mathbb N\), using the spatial conditional extremes model
\citep{wadsworth22_higher_spatial_extrem_singl_condit}. The model describes the conditional
distribution of \(X_t(\bm s)\) given that \(X_t(\bm s_0) > \tau_t(\bm s_0)\), where \(\bm s_0\) is
some chosen conditioning site and \(\tau_t(\bm s)\) is a large threshold that may vary in space and
time. The conditioning site can be located anywhere in \(\mathcal S\), which makes us free to place
\(\bm s_0\) at a specific location of interest, for modelling and simulating only the extremes that
we care about.

\begin{figure}
  \begin{center}
    \includegraphics[width=.95\linewidth]{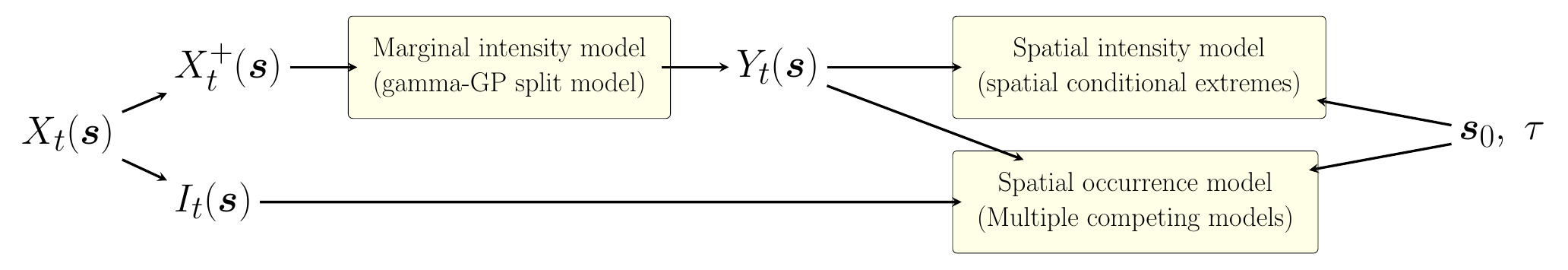}
  \end{center}
  \caption{%
    Flowchart visualising the complete modelling procedure. Precipitation $X_t(\bm s)$ is split
    into occurrence $I_t(\bm s)$ and intensity $X_t^+(\bm s)$.  The intensity marginals are modelled
    with a gamma-GP split-model (see Section~\ref{sec:marginal_model}), and $X_t^+(\bm s)$ is
    standardised into $Y_t(\bm s)$, with Laplace marginals. Then, given a conditioning site
    $\bm s_0$ and threshold $\tau$, the spatial conditional extremes model is fitted to the spatial
    intensity process, $[Y_t(\bm s) \mid Y_t(\bm s_0) > \tau]$ (see
    Section~\ref{sec:conditional_intensity_model}), and the spatial occurrence model is fitted to
    the occurrence process $[I_t(\bm s) \mid Y_t(\bm s_0) > \tau]$ (see
    Section~\ref{sec:conditional_occurrence_model}). Simulation from the model framework can be
    visualised by flipping most of the arrows in the flowchart, and thus moving from the right side
    to the left side.}
  \label{fig:flowchart}
\end{figure}

To use the spatial conditional extremes model, we must first transform \(X_t(\bm s)\) into a
standardised process \(\tilde X_t(\bm s)\), with Laplace marginal distributions at all locations and
time points in \(\mathcal S \times \mathcal T\), using the probability integral transform
\( \tilde X_t(\bm s) = F^{-1}[F_{\bm s, t}(X_t(\bm s))], \) where \(F^{-1}\) is the quantile
function of the Laplace distribution and \(F_{\bm s, t}\) is the cumulative distribution function of
\(X_t(\bm s)\) \citep{keef13_estim_condit_distr_multiv_variab}.  However, the marginal distribution
of hourly precipitation contains a point mass at zero, which means that the marginal distribution of
\(\tilde X_t(\bm s)\) also contains a point mass, making it different from the Laplace distribution
in its lower tail. \citet{richards22_model_extrem_spatial_aggreg_precip_condit_method,
  richards23_joint_extrem_spatial_aggreg_precip} tackle this problem by left-censoring all
zeros. This yields promising results, but the censoring makes high-dimensional inference
computationally intractable without the use of low-dimensional composite likelihoods. Moreover, this
approach is still computationally demanding as it relies on evaluating bivariate Gaussian
distributions many times. We therefore propose another approach for modelling \(X_t(\bm s)\) with
the spatial conditional extremes model.  Assume that hourly precipitation can be represented as
\( X_t(\bm s) = X^+_t(\bm s) I_t(\bm s), \) where
\(X^+_t(\bm s) = [X_t(\bm s) \mid X_t(\bm s) > 0]\) represents precipitation intensity, and
the binary random process \(I_t(\bm s)\), which equals \(1\) when \(X_t(\bm s) > 0\) and \(0\) when
\(X_t(\bm s) = 0\), represents precipitation occurrence.
So-called Richardson-type stochastic weather generators depend on this
formulation by first simulating the precipitation occurrence \(I_t(\bm s)\) and then simulating the
precipitation intensity \(X^+_t(\bm s)\) if \(I_t(\bm s) = 1\)
\citep{richardson81_stoch_simul_daily_precip_temper_solar_radiat}.  We build upon this approach by,
instead of modelling \([X_t(\bm s) \mid X_t(\bm s_0) > \tau_t(\bm s_0)]\), performing separate modelling
of the conditional intensity process \([X^+_t(\bm s) \mid X_t(\bm s_0) > \tau_t(\bm s_0)]\) and the
conditional occurrence process \([I_t(\bm s) \mid X_t(\bm s_0) > \tau_t(\bm s_0)]\), and then setting
\begin{equation}
  \label{eq:conditional_separation}
  \left[X_t(\bm s) \mid X_t(\bm s_0) > \tau_t(\bm s_0)\right] = \left[X^+_t(\bm s) I_t(\bm s) \mid
    X_t(\bm s_0) > \tau_t(\bm s_0)\right].
\end{equation}
The marginal distribution \(F^+_{\bm s, t}\) of \(X^+_t(\bm s)\) does not contain a point mass, so
it can more easily be transformed into the Laplace distribution. We denote the transformed intensity variables as \( Y_t(\bm s) = F^{-1}[F^+_{\bm s, t}(X^+_t(\bm s))]\). Thus, we describe the conditional
intensity process with the spatial conditional extremes model, while the conditional occurrence
process is described with a suitable binary model. Our model for \(F^+_{\bm s, t}\) is described in
Section~\ref{sec:marginal_model}. Then, our model for the conditional intensity process is described
in Section~\ref{sec:conditional_intensity_model}, and our model for the conditional occurrence
process is described in Section~\ref{sec:conditional_occurrence_model}. The full model structure is
visualised as a flowchart in Figure~\ref{fig:flowchart}. Most of our models fall within the
framework of latent Gaussian models, which are introduced in Section~\ref{sec:latent_gaussian}.

\subsection{Latent Gaussian Models}%
\label{sec:latent_gaussian}

A latent Gaussian model is a model where the observations \(\bm y = {(y_1, y_2, \ldots, y_n)}^\top\)
are assumed to be conditionally independent given a latent Gaussian random field
\(\bm x = {(x_1, x_2, \ldots, x_m)}^\top\) and a set of hyperparameters \(\bm \theta_1\), namely
\begin{equation}
  \left[ \bm y \mid \bm x, \bm \theta_1\right]
  \sim \prod_{i = 1}^n \pi(y_i \mid \eta_i(\bm x), \bm \theta_1), \qquad
  \left[ \bm x \mid \bm \theta_2 \right]
  \sim \mathcal N(\bm \mu(\bm \theta_2), \bm Q^{-1}(\bm \theta_2)),
\end{equation}
where the likelihood \(\prod_{i = 1}^n \pi(y_i \mid \eta_i(\bm x), \bm \theta_1)\) is a parametric
distribution with parameters \(\eta_i(\bm x)\) and \(\bm \theta_1\), the linear predictor
\(\eta_i(\bm x)\) is a linear combination of the elements in \(\bm x\) and the latent field
\(\bm x\) is conditionally Gaussian with mean vector \(\bm \mu(\bm \theta_2)\) and precision matrix \(\bm Q(\bm \theta_2)\),
given the hyperparameters \(\bm \theta_2\). The prior distributions of \(\bm \theta_1\) and
\(\bm \theta_2\) are \(\pi(\bm \theta_1)\) and \(\pi(\bm \theta_2)\), respectively.

The latent Gaussian modelling framework is highly flexible, as the likelihood can stem from an
essentially arbitrary parametric distribution, while information from explanatory variables and a
large variety of dependency structures can be incorporated into the linear predictor
\(\eta_i(\bm x)\). Additionally, non-Gaussian structures can be incorporated into the model through
the likelihood and the hyperparameters \(\bm \theta_1\) and \(\bm \theta_2\), which can be given any
kind of prior distributions.  Another advantage of the latent Gaussian modelling framework is that it
allows for fast approximate inference using INLA, which is implemented in the \texttt{R-INLA}
package \citep{niekerk21_new_front_bayes_model_using, niekerk23_new_avenue_bayes_infer_inla}. The
package contains a large range of pre-implemented model components for the linear predictor,
including splines, autoregressive (AR) models, random walk models and the so-called stochastic partial differential
equation (SPDE) model of \citet{lindgren11_explic_link_between_gauss_field}, which produces sparse
approximations of Gaussian random fields with Matérn autocorrelation function
\begin{equation}
  \label{eq:matern}
  \gamma(d) = \frac{1}{2^{\nu - 1}\Gamma(\nu)}{(\kappa d)}^\nu K_\nu(\kappa d),
\end{equation}
where \(d\) is the distance between two locations, \(\nu > 0\) is the smoothness parameter,
\(\rho = \sqrt{8 \nu} / \kappa\) is the range parameter and \(K_\nu\) is the modified Bessel
function of the second kind and order \(\nu\). Thus, \texttt{R-INLA} and the latent Gaussian modelling
framework make it easy to quickly develop and perform inference with complex models for a large
variety of applications.

\subsection{Modelling the marginals}%
\label{sec:marginal_model}

We model marginal distributions of the intensity process \(X^+_t(\bm s)\) by first modelling the
bulk of the data with the gamma distribution, and then the upper tail with the GP
distribution. Specifically, we model margins as
\begin{equation}
  \label{eq:marginals}
  F^+_{\bm s, t}(x) =
  \begin{cases}
    G_{\bm s, t}(x) & x \leqslant u_t(\bm s), \\
    G_{\bm s, t}(u_t(\bm s)) + (1 - G_{\bm s, t}(u_t(\bm s))) H_{\bm s, t}(x -
    u_t(\bm s)) & x > u_t(\bm s),
  \end{cases}
\end{equation}
where \(G_{\bm s, t}\) and \(H_{\bm s, t}\) are cumulative distribution functions of the gamma and
GP distributions respectively, both with parameters that might vary in space and time, while
\(u_t(\bm s)\) is the \(p_u\)-quantile of \(G_{\bm s, t}\), for some large probability \(p_u\).
This ``split-modelling'' approach is a common choice for modelling precipitation when aiming to
describe both the bulk and the upper tail of the distribution
\citep[e.g.,][]{CameronEtAl2000Modellingextremerainfalls, VracEtAl2007Stochasticdownscalingprecipitation, opitz18_inla_goes_extrem}. 
We use the gamma and GP parametrisations of
\citet{castro-camilo19_splic_gamma_gener_paret_model}, which give a probability density function for
the gamma distribution of the form
\begin{equation}
  g(x) = \frac{x^{\kappa - 1}}{\Gamma(\kappa)}
  {\left(\frac{G^{-1}(\alpha; \kappa, 1)}{\psi_\alpha}\right)}^\kappa \exp{\left(-x
      \frac{G^{-1}(\alpha; \kappa, 1)}{\psi_\alpha}\right)},\quad x > 0, \quad \kappa, \psi_\alpha > 0,
\end{equation}
where \(\kappa\) is the standard shape parameter, \(\alpha \in (0, 1)\) is a fixed probability,
\(G^{-1}(\alpha; \kappa, 1)\) is the \(\alpha\)-quantile of a gamma distribution with shape
\(\kappa\) and scale \(1\), and the parameter \(\psi_\alpha\) is equal to the \(\alpha\)-quantile of
\(g(x)\).  Using this parametrisation for the likelihood of our latent Gaussian model lets us
directly estimate the \(\alpha\)-quantile of the data through the parameter \(\psi_\alpha\). The GP
distribution has cumulative distribution function
\begin{equation}
  H(x) =
  \begin{cases}
    1 - {\left(1 + \left\{{(1 - \beta)}^{-\xi} - 1\right\} \frac{x}{\phi_\beta}\right)}_{+}^{-1 / \xi},
    & \xi \neq 0, \\
    1 - {(1 - \beta)}^{x / \phi_\beta},
    & \xi = 0,
  \end{cases} \quad \phi_\beta > 0, \xi \in \mathbb R,
\end{equation}
with \({(a)}_{+} = \max(0, a)\), where \(\xi\) is the tail parameter of the GP distribution, \(\beta\)
is a fixed probability and the parameter \(\phi_\beta\) equals the \(\beta\)-quantile of the GP
distribution. The support of the GP distribution is \((0, \infty)\) for \(\xi \geqslant 0\), while
it is \(\left(0, \phi_\beta \left\{1 - {(1 - \beta)}^{-\xi}\right\}\right)\) for \(\xi < 0\).

We estimate the parameters of \(G_{\bm s, t}\) and \(H_{\bm s, t}\) separately, with two different
latent Gaussian models. For the parameters of \(G_{\bm s, t}\), we use a latent Gaussian model with
a gamma likelihood, where the shape parameter \(\kappa\) is a hyperparameter that is constant in
space and time, while \(\psi_\alpha\) is allowed to vary in space and time through the linear
predictor \(\eta = \log \psi_\alpha\). Setting \(\alpha = p_u\) lets us directly estimate the
threshold \(u_t(\bm s)\), while estimating the parameters of \(G_{\bm s, t}\). The value of \(p_u\),
the components of \(\eta\) and the priors for \(\bm \theta\) vary depending on the application, and
are therefore described in Section~\ref{sec:case_study:marginal_model}. After having performed
inference with \texttt{R-INLA}, we estimate the parameters of \(G_{\bm s, t}\) as the posterior
means of \(\kappa\) and \(\psi_\alpha\).

Having estimated the threshold \(u_t(\bm s)\), we then model the distribution \(H_{\bm s, t}\) of
the threshold exceedances \([X^+_t(\bm s) - u_t(\bm s) \mid X^+_t(\bm s) > u_t(\bm s)]\) with the GP
distribution. Here, we apply a latent Gaussian model with a GP likelihood, where the tail parameter
\(\xi\) is a hyperparameter that is constant in space and time, while the linear predictor is
\(\eta = \log \phi_\beta\), where we set \(\beta = 0.5\), so that \(\phi_\beta\) is the GP
median, allowed to vary in space and time. Once more, the parameters of \(H_{\bm s, t}\) are estimated as the posterior means of
\(\xi\) and \(\phi_\beta\). Note that the GP likelihood within \texttt{R-INLA} only allows for
modelling \(\xi > 0\). However, this should not be too problematic when modelling hourly
precipitation data, as there is considerable evidence in the literature that precipitation is
heavy-tailed, and thus it should be modelled with a non-negative tail parameter, especially for
short temporal aggregation times \citep{ cooley07_bayes_spatial_model_extrem_precip_retur_level,
  van12_spatial_regres_model_extrem_precip_belgium, papalexiou13_battl_extrem_value_distr,
  huser14_space_time_model_extrem_event}.

\subsection{Spatial modelling of the conditional intensity process}%
\label{sec:conditional_intensity_model}

We transform the precipitation intensities \(X^+_t(\bm s)\) into the standardised process
\(Y_t(\bm s)\), with Laplace marginal distributions, using the probability integral transform,
\(Y_t(\bm s) = F^{-1}[F^+_{\bm s, t}(X^+_t(\bm s))]\). Given a conditioning site \(\bm s_0\) and a
threshold \(\tau_t(\bm s_0)\), we then model the spatial distribution of \(Y_t(\bm s)\) given that
\(X_t(\bm s_0) > \tau_t(\bm s_0)\), which is the same conditioning event as
\(Y_t(\bm s_0) > F^{-1}[F^+_{\bm s_0, t}(\tau_t(\bm s_0))]\). We assume that the notion of
``extremes'' can vary across time and space on the original precipitation scale, but not on the
transformed Laplace scale. We therefore set \(\tau_t(\bm s_0)\) equal to a chosen quantile of
\(F^+_{\bm s_0, t}\), which gives the constant threshold
\(\tau = F^{-1}[F^+_{\bm s_0, t}(\tau_t(\bm s_0))]\) on the Laplace scale. The spatial conditional
extremes model of \citet{SimpsonEtAl2023Highdimensionalmodeling} now states that, for \(\tau\)
large enough, the conditional process \(\left[Y_t(\bm s) \mid Y_t(\bm s_0) = y_0 > \tau\right]\) is
Gaussian, with the structure
\begin{equation}
  \label{eq:intensity_model}
  \left[Y_t(\bm s) \mid Y_t(\bm s_0) = y_0 > \tau\right] \overset{d}{=} a(\bm s; \bm s_0,
  y_0) + b(\bm s; \bm s_0, y_0) Z_t(\bm s; \bm s_0) + \varepsilon_t(\bm s; \bm s_0),
\end{equation}
where \(a(\cdot)\) and \(b(\cdot)\) are two suitable standardising functions such that \(a(\bm s_0; \bm s_0, y_0) = y_0\) and \(\lvert a(\bm s; \bm s_0, y_0)\rvert \leq y_0\), \(b(\bm s; \bm s_0, y_0) > 0\) for all \(\bm s \in \mathcal S\), \(Z_t(\bm s; \bm s_0)\) is a
spatial Gaussian random field with \(Z_t(\bm s_0; \bm s_0) = 0\) almost surely, and
\(\varepsilon_t(\bm s; \bm s_0)\) is Gaussian white noise with
\(\varepsilon_t(\bm s_0; \bm s_0) = 0\) almost surely. This is the same as a latent Gaussian model
with a Gaussian likelihood and latent field \(a(\cdot) + b(\cdot) Z(\cdot)\), which means that
computationally efficient approximate inference can be performed using
INLA\@. \citet{SimpsonEtAl2023Highdimensionalmodeling} demonstrate how to perform efficient
high-dimensional inference by using \texttt{R-INLA} and modelling \(Z_t(\bm s; \bm s_0)\) with the
SPDE approximation.  \citet{VandeskogEtAl2024efficientworkflowmodelling} build upon
their work and develop a methodology for implementing computationally efficient parametric models
for \(a(\bm s; \bm s_0, y_0)\) and \(b(\bm s; \bm s_0, y_0)\) in \texttt{R-INLA} and a method for efficient
constraining of \(Z_t(\bm s; \bm s_0)\) such that \(Z_t(\bm s_0; \bm s_0) = 0\) almost surely. We
apply their methodology for modelling the spatial distribution of conditional precipitation
extremes, while developing new diagnostics and models for the standardising functions \(a(\cdot)\)
and \(b(\cdot)\).

\subsection{Spatial modelling of the conditional occurrence process}%
\label{sec:conditional_occurrence_model}

Four competing models are applied to describe the spatial distribution of conditional precipitation
occurrences, \(\left[I_t(\bm s) \mid X_t(\bm s_0) > \tau_t(\bm s_0)\right] \equiv I_t(\bm s; \bm s_0) \). One of these is the
relatively common spatial probit model, which assumes that \(I_t(\bm s; \bm s_0)\) depends on an underlying
latent Gaussian process \(\mathcal W_t(\bm s; \bm s_0)\) such that \(I_t(\bm s; \bm s_0) = 1\) when
\(\mathcal W_t(\bm s; \bm s_0) + \varepsilon_t(\bm s) \geqslant 0\) and \(I_t(\bm s; \bm s_0) = 0\) otherwise, where
\(\varepsilon_t(\bm s)\) is zero-mean Gaussian white noise with a fixed variance of 1.
This Gaussian nugget effect makes the process \(I_t(\bm s; \bm s_0)\) 
conditionally independent given \(\mathcal W_t(\bm s; \bm s_0)\), with distribution
\(\text{P}\left[I_t(\bm s; \bm s_0) = 1 \mid \mathcal W_t(\bm s; \bm s_0)\right] = \Phi(\mathcal W_t(\bm s; \bm s_0))\), where
\(\Phi(\cdot)\) is the cumulative distribution function of the standard normal distribution. This means that the probit model is in
fact a latent Gaussian model with a Bernoulli likelihood, that is amenable to fast inference
using INLA\@. The nugget variance is fixed to 1 to ensure model identifiability. Within \texttt{R-INLA}, we decompose \(\mathcal W_t(\bm s; \bm s_0)\) into
\(\mathcal W_t(\bm s; \bm s_0) = \mu(\bm s; \bm s_0) + \mathcal W_t^*(\bm s; \bm s_0)\). Here,
\(\mu(\bm s; \bm s_0)\) is the mean value of \(\mathcal W_t(\bm s; \bm s_0)\), 
which therefore captures the overall probability that precipitation occurs a certain distance away from a storm center.
The random field \(\mathcal W_t^*(\bm s; \bm s_0)\) is a zero-mean Gaussian random
field that captures all the time-dependent fluctuations of \(\mathcal W_t(\bm s; \bm s_0)\).
For faster inference, we model
\(\mathcal W_t^*(\bm s; \bm s_0)\) with the SPDE approximation. We enforce \(I_t(\bm s_0; \bm s_0) = 1\)
by modelling \(\mu(\bm s; \bm s_0)\) as a function of the distance \(d\) between \(\bm s\) and \(\bm s_0\),
i.e., \(\mu(\bm s; \bm s_0) \equiv \mu(d)\) with \(d = \|\bm s - \bm s_0\|\). Then we ensure that \(\mu(0)\)
is positive and large, while also enforcing that \(\mathcal W_t^*(\bm s_0; \bm s_0) = 0\) almost surely, using the
constraining method of \citet{VandeskogEtAl2024efficientworkflowmodelling}. This does
not guarantee that P\((I_t(\bm s_0; \bm s_0) = 1) = 1\) exactly, but if \(\mu(0)\) is large enough, then
P\((I_t(\bm s_0; \bm s_0) = 1) \approx 1\) for most practical purposes. The exact structure of \(\mu(d)\)
varies depending on the application in question.

The spatial probit model can produce realistic realisations of the spatial binary process, but it can
also struggle in situations where the binary field is smooth, in the sense that the variance of
\(\varepsilon_t(\bm s)\) is considerably smaller than the variance of \(\mathcal W_t(\bm s; \bm s_0)\). To ensure
smooth model realisations, the variance of \(\mathcal W_t(\bm s; \bm s_0)\) must become so large that the
probability \(\Phi(\mathcal W_t(\bm s; \bm s_0))\) always is close to either 0 or 1, and this large variance
can make it difficult to reliably estimate trends in the mean \(\mu(d)\). For this reason, we
also attempt to model the conditional occurrence process using a probit model without any spatial
effects, i.e., where we remove \(\mathcal W_t^*(\bm s; \bm s_0)\). This model typically fails at providing
realistic-looking realisations of smooth binary processes, but it can perform considerably better at
capturing trends in the mean structure.

Our probit models are independent of the conditional intensity model, so it is possible for the
simulated occurrence samples to create highly non-smooth precipitation realisations where areas with
large precipitation values suddenly contain a ``hole'' of zeros close to the most extreme
observations. This is an unrealistic behaviour that we wish to avoid. Our third modelling strategy
is therefore based on the assumption that the occurrence process is dependent upon the intensity
process such that only the smallest values of \(X_t^+(\bm s)\) get turned into zeros. Thus, the
third approach, denoted the threshold model, estimates the overall probability \(p\) of observing
zeros in the data, and then sets \(I_t(\bm s; \bm s_0)\) equal to zero whenever \(X_t^*(\bm s)\) is smaller
than its estimated \(p\)-quantile. Lastly, for improved baseline comparisons, we add a fourth
occurrence model, which interprets ``no precipitation'' as a tiny but positive amount of
precipitation, i.e., \(I_t(\bm s; \bm s_0) \equiv 1\) for all \(\bm s, \bm s_0 \in \mathcal S\). We denote this the nonzero model.

\section{Simulating extreme hourly precipitation}%
\label{sec:case_study}

We apply the models from Section~\ref{sec:model_framework} to the data from
Section~\ref{sec:data} for modelling and simulating spatial realisations of extreme hourly
precipitation. In Section~\ref{sec:case_study:marginal_model} we model and standardise the marginal
distributions of the precipitation radar data. Then, in Section~\ref{sec:case_study:intensity} and
Section~\ref{sec:case_study:occurrence}, we model the conditional intensities and occurrences of
extreme precipitation, respectively. Finally, in Section~\ref{sec:case_study:final_simulations}, we
combine all the model fits for simulating spatial realisations of extreme hourly precipitation.

\subsection{Marginals}%
\label{sec:case_study:marginal_model}

We model the marginal distribution \(F^+_{\bm s, t}\) of nonzero radar precipitation data using the
gamma-GP split-model from Section~\ref{sec:marginal_model}, where we choose \(\alpha = p_u = 0.95\).
Other values of \(p_u\) were tested, but they did not produce considerably different results.
We first attempt to model the linear predictor with a separable space-time model, more specifically as the sum of a Gaussian smoothing spline depending on elevation, a Gaussian smoothing spline depending on time and a spatial Gaussian random field.
Here, both the spatial Gaussian random field and the two Gaussian smoothing splines are modelled using the SPDE approximation.
However, we found that both the spatial and elevation effects were estimated to be nearly constant. As a result, a simpler temporal model, using only the Gaussian temporal smoothing spline, provided a fit that was just as good as the more complex model, while allowing for  more robust and computationally efficient inference. Therefore, we opted to use the purely temporal model, which includes only the temporal Gaussian smoothing spline in the linear predictor.
A model based on splines is unsuitable for prediction outside the observed
spatio-temporal domain, but the aim of this paper is modelling, rather than forecasting, so we find it
to be a good model choice.

\begin{figure}
  \centering
  \includegraphics[width=.9\linewidth,page=3]{marginal_bulk.pdf}
  \includegraphics[width=.9\linewidth,page=1]{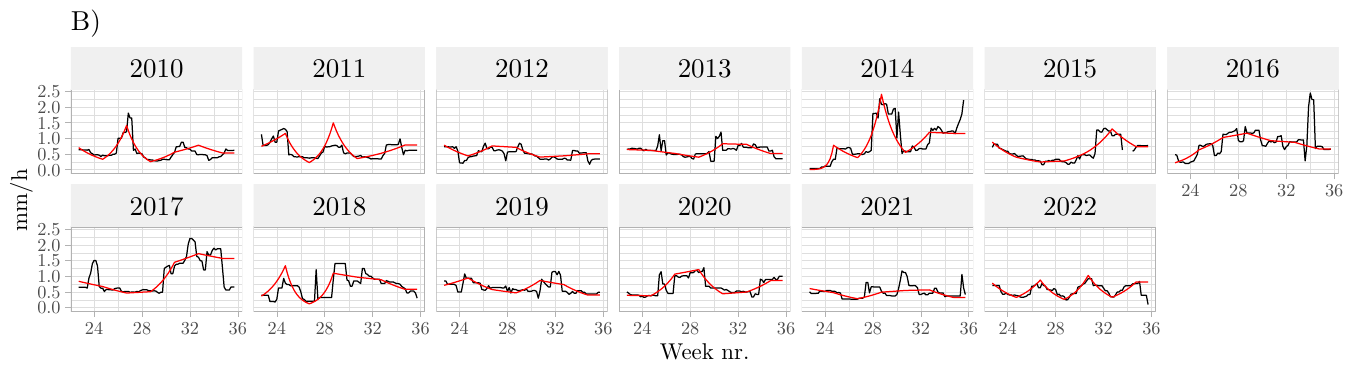}
  \caption{Estimates of the \(95\%\)-quantile of hourly precipitation intensity (A) and the median
    of precipitation threshold exceedances (B) for each summer day during 2010--2022. In black:
    empirical estimates created using a sliding window of width of one week. In red: posterior
    means estimated using the latent Gaussian models with a gamma likelihood (A) and a GP likelihood
    (B).}%
  \label{fig:marginal_bulk}
\end{figure}

The penalised complexity (PC) prior is a weakly informative prior that allows one to incorporate prior knowledge, while also being
designed to penalise model complexity, by placing an exponential prior on the Kullback--Leibler divergence from some simpler base model \citep{simpson17_penal_model_compon_compl}.
We place the PC prior of 
\citet{fuglstad19_const_prior_that_penal_compl} on the
range \(\rho = \sqrt{8 \nu} / \kappa\) and variance \(\sigma^2\) such that the prior probability
that \(\rho\) exceeds \(28\)~days is \(5\%\) and the prior probability that \(\sigma\) exceeds a
value of \(3\) is \(5\%\).
The smoothness parameter \(\nu\) can be difficult to estimate
\citep{lindgren15_bayes_spatial_model_r_inla}, so we fix it to \(\nu = 1.5\).
This adequately represents our assumption that the temporal spline should be relatively smooth, 
as we do not expect abrupt or significant changes in the marginal distributions between precipitation observations that are only a week apart.
The value of \(1.5\) is also the default and the most extensively tested value within \texttt{R-INLA}~\citep{lindgren15_bayes_spatial_model_r_inla}.
Note, however, that even though the choice of \(\nu\) can heavily affect the posterior distributions of \(\rho\) and \(\sigma\),
the considerable flexibility of the Gaussian random field model
causes the posterior mean of the random field itself at each time point and location to be almost unaffected by the choice of \(\nu\).
Thus, we also tested other values of \(\nu\), but we found no discernible differences in the posterior means of \(\kappa\) and \(\psi_\alpha\).

Due to the large
amounts of data, we speed up inference for the gamma model by only using observations from a spatial subgrid of the data
with resolution \(2 \times 2\)~km\(^2\).
Inference is performed with \texttt{R-INLA}, in approximately half an hour, when using only one core
on a 2.6~gHz Linux server with approximately 200 GB RAM. The posterior mean of the shape parameter \(\kappa\) equals \(0.69\), and
the posterior mean of the threshold \(u_t(\bm s)\) is displayed in Subplot A of
Figure~\ref{fig:marginal_bulk}, together with the empirical \(95\%\)-quantiles of the precipitation
radar intensities, pooled across space. The smoothing spline seems to capture the temporal trends of
the data well. We evaluate the model fit using quantile-quantile (QQ) plots. These are displayed in
Figures~\ref{fig:qq} and~\ref{fig:qq_gamma_random} of the Supplementary Material, and they show an
almost perfect correspondence between model quantiles and empirical quantiles. We conclude that the
model fit is satisfactory.

Having estimated \(u_t(\bm s)\), we then model the threshold exceedances
\([X^+_t(\bm s) - u_t(\bm s) \mid X^+_t(\bm s) > u_t(\bm s)]\) with the GP distribution, as
described in Section~\ref{sec:marginal_model}. Once more, we find that a purely temporal 
linear predictor performs just as well as a separable space-time model, while  allowing for more robust and computationally efficient inference.
Therefore, we use a temporal spline for the linear predictor, similar to the one employed in the gamma model, with the same prior distributions.
The tail parameter \(\xi\) is given the PC prior of \citet{opitz18_inla_goes_extrem},
\begin{equation}
  \pi(\xi) = \lambda (1 - \xi / 2) {(1 - \xi)}^{-3/2} 2^{-1/2} \exp \left(-\lambda \xi / \sqrt{2 (1
      - \xi)}\right), \quad 0 \leqslant \xi < 1,
\end{equation}
with penalty parameter \(\lambda\). The GP distribution has infinite mean for \(\xi \geqslant 1\)
and infinite variance for \(\xi \geqslant 1/2\), and since it is well established in the literature
that \(\xi\) tends to be in the range between \(0.05\) and \(0.3\) for precipitation data
\citep[e.g.,][]{cooley07_bayes_spatial_model_extrem_precip_retur_level,
  van12_spatial_regres_model_extrem_precip_belgium, papalexiou13_battl_extrem_value_distr}, we
enforce \(\xi \leqslant 1 / 2\) to ease parameter estimation. Then, we choose \(\lambda = 7\), which
gives the prior probability P\((\xi \leqslant 0.4) \approx 95\%\).

Inference is performed with \texttt{R-INLA}, using all the spatial locations, in approximately
2~minutes. The posterior mean of \(\xi\) is \(0.145\), which is far away from the upper bound of
\(1/2\), and matches well the results of \citet{vandeskog22_model_sub_precip_extrem_blend},
who estimated \(\xi = 0.18\) with a \(95\%\) credible interval of \((0.14, 0.21)\) when modelling
the yearly maxima of hourly precipitation using rain gauge data from a spatial domain that covers
\(\mathcal S\).  Subplot B of Figure~\ref{fig:marginal_bulk} displays the empirical median of the
threshold exceedances, pooled across space, along with the posterior means of the threshold exceedance
medians, which seem to agree with the main temporal trends of the data well. We evaluate the model
fit using QQ plots, displayed in Figures~\ref{fig:qq} and~\ref{fig:qq_gp_random} in the Supplementary Material. These demonstrate a good correspondence
between model quantiles and empirical quantiles overall, although with a slight underestimation of quantiles larger than the $1 - 1/(9 \times 92 \times 24)\approx0.99995$-quantile of hourly summer precipitation data.
In the Supplementary Material we discuss some of the potential issues with the marginal model that might lead to this underestimation, and how they might be improved upon in future research. 
However, in any case, the marginal fit is quite reasonable, and appropriate at all levels that we consider in the data application discussed in this paper.

The marginal model fit is further evaluated by comparing empirical moments of the observations with moments of the fitted marginal distributions,
as displayed in Figures~\ref{fig:pit} and~\ref{fig:gp_pit} of the Supplementary Material.
Even though some minor trends in bias and under/overdispersion are present in these figures, we once more conclude that our model provides a
satisfactory fit to the data for this paper.
 
\subsection{Conditional intensity process}%
\label{sec:case_study:intensity}

We standardise the precipitation intensities to have Laplace marginal distributions.  Then,
following \citet{keef13_estim_condit_distr_multiv_variab}, we choose functions of the form
\(a(\bm s; \bm s_0, y_0) = \alpha(\bm s; \bm s_0) y_0\) and
\(b(\bm s; \bm s_0, y_0) = y_0^{\beta(\bm s; \bm s_0)}\) for the spatial conditional extremes
model~\eqref{eq:intensity_model}, which, they claim, can cover a large range of dependence
structures, including all the standard copulas studied by
\citet{joe97_multiv_model_multiv_depen_concep} and \citet{nelsen06_introd_copul}. Building upon the
work of \citet{VandeskogEtAl2024efficientworkflowmodelling}, we develop new empirical
diagnostics for making informed decisions about the value of the threshold \(\tau\) and the forms of
\(\alpha(\bm s; \bm s_0)\) and \(\beta(\bm s; \bm s_0)\), taking values in \([-1, 1]\) and \([0, 1]\), respectively.

We assume that the standardising functions only depend on the Euclidean distance to \(\bm s_0\) and
define \(\alpha(d) \equiv \alpha(\bm s; \bm s_0)\) with \(d = \|\bm s - \bm s_0\|\), and similarly
for \(\beta(d)\). Assuming that the residual field \(Z_t(\bm s; \bm s_0)\) is isotropic, we denote
the mean and variance of \([Y_t(\bm s) \mid Y_t(\bm s_0) = y_0]\) as \(\mu(d; y_0)\) and
\({\zeta(d; y_0)}^2\), respectively. Under the spatial conditional extremes
model~\eqref{eq:intensity_model}, these equal \(\mu(d; y_0) = \alpha(d) y_0\) and
\({\zeta(d; y_0)}^2 = {\sigma(d)}^2 y_0^{2 \beta(d)} + \sigma_\varepsilon^2\), where
\(\sigma_\varepsilon^2\) is the variance of the nugget term \(\varepsilon_t(\bm s; \bm s_0)\) and
\({\sigma(d)}^2\) is the variance of \(Z_t(\bm s; \bm s_0)\) when \(\|\bm s - \bm s_0\| = d\).  By
computing the empirical mean, \(\hat \mu(d; y_0)\), of the conditional precipitation intensity, we
can estimate \(\alpha(d)\) as \(\hat \alpha(d; y_0) = \hat \mu(d; y_0) / y_0\).  Similarly, by
assuming that \(\sigma_\varepsilon\) is small, we can estimate \(\beta(d)\) using empirical
conditional variances as
\begin{equation}
  \hat \beta(d; y_0, y_1) = \log\{\hat \zeta(d; y_0) / \hat \zeta(d; y_1)\} / \log(y_0 / y_1), 
\end{equation}
where \(\hat \zeta(d; y)\) is the empirical standard deviation of all observations at distance
\(d\) from a conditioning site with threshold exceedance \(y\), and \(y_0 \neq y_1\) are any two
threshold exceedances \(y_0, y_1 > \tau\). This allows us to create multiple different
estimates \(\hat \alpha(d; y_0)\) and \(\hat \beta(d; y_0, y_1)\) for \(\alpha(d)\) and
\(\beta(d)\), respectively, by varying the values of \(y_0\) and \(y_1\). This provides a diagnostic
for estimating the threshold \(\tau\), since the spatial conditional extremes model assumes that
\(\alpha(d)\) and \(\beta(d)\) are constant for all threshold exceedances larger than
\(\tau\). Thus, we compute \(\hat \alpha(d; y_0)\) and \(\hat \beta(d; y_0, y_1)\) for a large range
of values of \(y_0\) and \(y_1\), and we set \(\tau\) equal to the smallest value such that the
estimates are approximately constant for all \(y_0, y_1 > \tau\). Then, we propose parametric
functions for \(\alpha(d)\) and \(\beta(d)\) that can fit well to the patterns that we find in the
empirical estimates. Finally, we also compute
\(\hat \sigma(d; y_0, y_1) = \hat \zeta(d; y_0) y_0^{-\hat \beta(d; y_0, y_1)}\) to get an idea
about the marginal variance of the residual process \(Z(\bm s; \bm s_0)\).

\begin{figure}
  \centering
  \includegraphics[width=.99\linewidth,page=2]{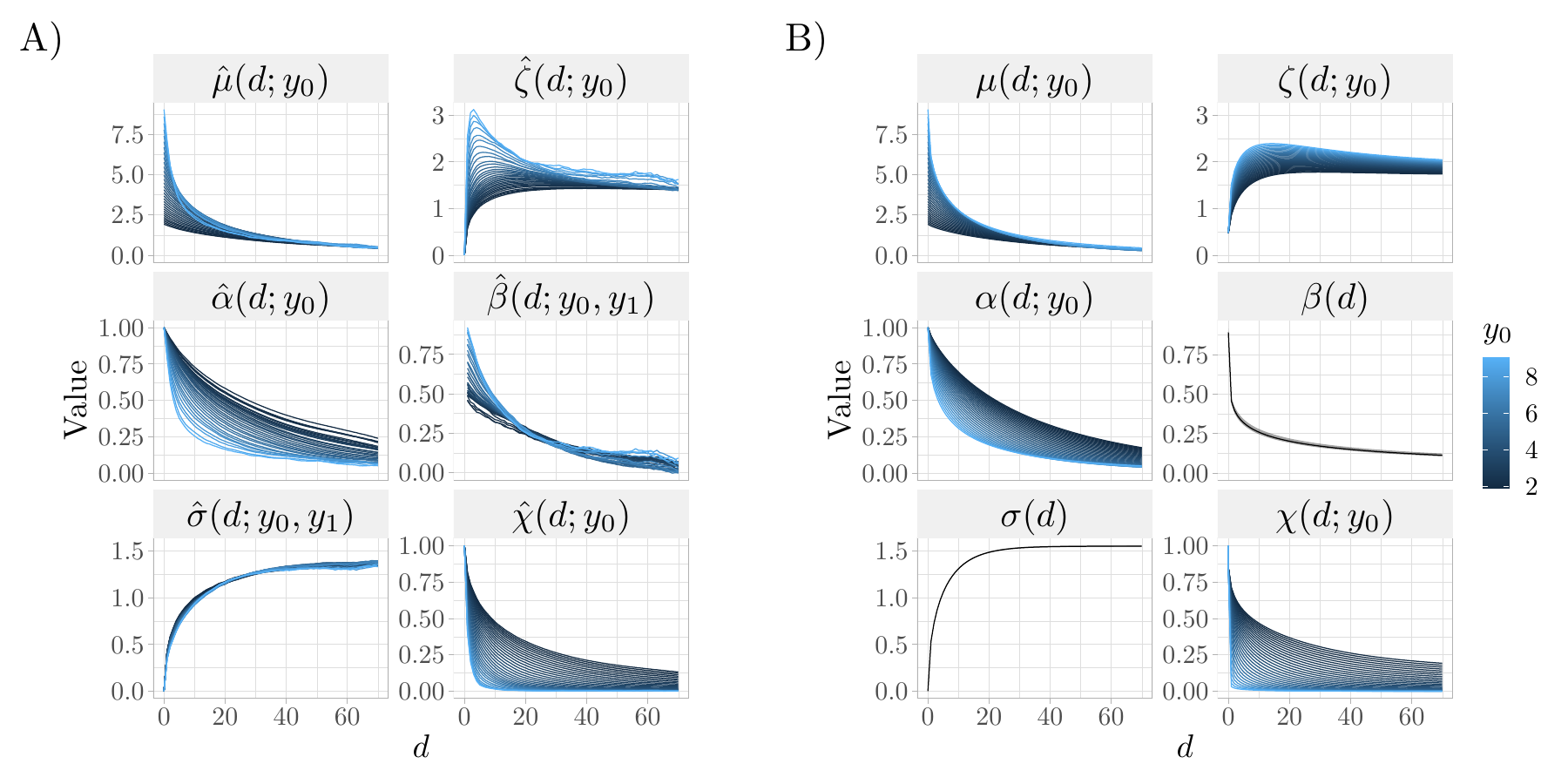}
  \caption{A) Empirical moments and estimates of \(\chi(d; y_0)\) and components of the
    conditional extremes model~\eqref{eq:intensity_model}.
    B) Posterior means of the same variables from the model fit based on conditioning site nr.~2.}%
  \label{fig:empirical_conditional}
\end{figure}
Exploratory analysis hints at some weak anisotropy in the precipitation data. However, we do not
believe the lack of isotropy is strong enough to cause considerable problems, and the development of
a suitable anisotropic model is outside the scope of this paper. We therefore assume an isotropic
model. We compute \(\hat \mu(d; y_0)\) and \(\hat \zeta(d; y_0)\) with a sliding window
approach. For any value of \(d\) and \(y_0\), the moments are estimated using all observations
within a distance \(d \pm 0.5\)~km from a location where a value of \(\log (y_0) \pm 0.025\) is
observed. We then compute \(\hat \alpha(d; y_0)\), \(\hat \beta(d; y_0, y_1)\) and
\(\hat \sigma(d; y_0, y_1)\) as previously described, where we fix \(y_1\) to the \(90\%\)-quantile
of the Laplace distribution. We also estimate empirical conditional exceedance probabilities
\(\chi(d; y_0) \equiv \chi_p(\bm s, \bm s_0)\), where \(d = \|\bm s - \bm s_0\|\) and
\(p = F^{-1}(y_0)\) with \(F^{-1}\) the quantile function of the Laplace distribution, using a
similar sliding window approach. Estimates created using \(d \in [1, 70]\)km and \(\log (y_0) \in [0.5, 2.2]\) are displayed in Subplot A of
Figure~\ref{fig:empirical_conditional}. The estimated shape of \(\sigma(d)\) matches well 
the standard deviation of a random field with constant variance, constrained to be zero at
\(\bm s_0\). The estimates of \(\alpha(d)\) seem to behave like functions with exponential-like
decay towards zero as \(d\) increases. However, the decay occurs at an increasing rate as \(y_0\)
increases, and it never seems to stabilise as a function of \(y_0\). This indicates that, with the
available amounts of data, we cannot choose a large enough threshold \(\tau\) such that the function
\(a(d; y_0) = \alpha(d) y_0\) provides a good fit for all \(y_0 > \tau\). Therefore, we instead
propose to model the mean as \(a(d; y_0) = \alpha(d; y_0) y_0\), where the function
\(\alpha(\cdot)\) depends on both distance \(d\) and intensity level \(y_0\), and we choose a
relatively low threshold, \(\tau = y_1\).

A common model for \(\alpha(d)\) is \(\alpha(d) = \exp\{-{(d / \lambda_a)}^{\kappa_a}\} \in [0, 1]\), with
\(\lambda_a, \kappa_a > 0\), see, e.g, \citet{wadsworth22_higher_spatial_extrem_singl_condit},
\citet{richards22_model_extrem_spatial_aggreg_precip_condit_method} and
\citet{simpson21_condit_model_spatio_tempor_extrem}. We therefore examine if the model
\(\alpha(d; y_0) = \exp[-{\{d / \lambda_a(y_0)\}}^{\kappa_a(y_0)}]\) can provide a good fit to our
data, where \(\lambda_a(y_0)\) and \(\kappa_a(y_0)\) are parametric functions of \(y_0\). Using a
sliding window approach, we estimate \(\lambda_a(y_0)\) and \(\kappa_a(y_0)\) by minimising the sum
of squared errors between \([Y(\bm s) \mid Y(\bm s_0) = y_0]\) and \(a(\|\bm s - \bm s_0\|; y_0)\)
for different values of \(y_0\). These least squares estimates are displayed in Figure~\ref{fig:a_mse_fit} of the Supplementary
Material. Based on our findings we propose to model \(\lambda_a(y_0)\) as
\(\lambda_a(y_0) = \lambda_{a0} \exp(-(y_0 - \tau) / \Lambda_\lambda)\), with
\(\lambda_{a0}, \Lambda_\lambda > 0\), and \(\kappa_a(y_0)\) as
\(\kappa_a(y_0) = \kappa_{a0} \exp(-{((y_0 - \tau) / \Lambda_\kappa)}^{\varkappa})\), with
\(\kappa_{a0}, \Lambda_\kappa, \varkappa > 0\), where \(y_0 > \tau\).

The estimates of \(\beta(d)\) in Subplot A of Figure~\ref{fig:empirical_conditional} have a clear
dependence on \(y_0\) at short distances \(d\). However, \(\hat \beta(d; y_0, y_1)\) seems to be
independent of \(y_0\) for longer distances, and the changes as a function of \(y_0\) are much less
severe than those in \(\hat \alpha(d; y_0)\). We therefore stick to a model on the form
\(b(d; y_0) = y_0^{\beta(d)}\). Based on the estimates in Subplot A of
Figure~\ref{fig:empirical_conditional}, it seems that \(\beta(d)\) should be modelled with a
function that decays exponentially towards zero. We therefore choose the model
 \(\beta(d) = \beta_0 \exp(-{(d / \lambda_b)}^{\kappa_b}) \in [0, \beta_0]\) with
\(\lambda_b, \kappa_b > 0\) and \(\beta_0 \in [0, 1)\).

Having chosen the threshold \(\tau\) and parametric forms for the functions \(a(d; y_0)\) and
\(b(d; y_0)\), we then apply the method of
\citet{VandeskogEtAl2024efficientworkflowmodelling} for defining a nonstationary and
constrained SPDE approximation to the spatial Gaussian random field in~\eqref{eq:intensity_model}
within \texttt{R-INLA}. This SPDE model approximates spatial Gaussian random fields of
the form \(b(\bm s; \cdot) Z(\bm s)\) as a linear combination of \(m\) Gaussian mesh nodes,
\(\hat Z_b(\bm s) = \sum_{i = 1}^m \phi_i(\bm s) b_i W_i\), where \(b_1, b_2, \ldots, b_m\) are the
values of the function \(b(\bm s)\) at the location of the \(m\) mesh nodes, and \(\phi_i\) and
\(W_i\) are basis functions and Gaussian mesh nodes from the ``standard'' SPDE approximation,
\(\hat Z(\bm s) = \sum_{i = 1}^m \phi_i(\bm s) W_i\), of
\citet{lindgren11_explic_link_between_gauss_field}. The nonstationary SPDE approximation is then
constrained at \(\bm s_0\) by placing one of the mesh nodes at \(\bm s_0\), and constraining it to
be exactly zero.

For each of the five conditioning sites displayed in Figure~\ref{fig:data_domain}, we perform inference with \texttt{R-INLA}, using
data from all time points where \(\tau\) is exceeded at \(\bm s_0\) and all 6404 locations in
\(\mathcal S\). In an empirical-Bayes-like approach, we place weakly informative Gaussian priors on the logarithms of
the parameters in \(a(\cdot)\), with variance \(5^2\) and means equal to their least squares
estimates.
The variance of \(5^2\) is chosen to make the prior distributions less informative, as sharp priors centered around the least squares estimates would carry a risk of overfitting by ``using the data twice''. 
For the parameters of \(b(\cdot)\), we similarly choose Gaussian priors with variance \(5^2\) for
\(\log (\lambda_b)\), \(\log (\kappa_b)\) and \(\log(\beta_0 / (1 - \beta_0))\), which ensures
\(\lambda_b, \kappa_b > 0\) and \(\beta_0 \in (0, 1)\). The prior means are chosen based on the
diagnostics in Figure~\ref{fig:empirical_conditional}. We set them equal to \(\log(8.5)\),
\(\log(0.5)\) and \(\log(0.65 / (1 - 0.65))\), respectively. The parameters of \(Z_t(\bm s; \bm s_0)\) are
given the PC prior of \citet{fuglstad19_const_prior_that_penal_compl}, such that the prior
probability that the range \(\rho\) exceeds \(60\)~km is \(5\%\) and the prior probability that the
standard deviation \(\sigma\) exceeds \(4\) is \(5\%\). We fix the smoothness parameter to
\(\nu = 0.5\), to represent our belief about the smoothness properties of extreme precipitation
fields. This is a considerably smaller smoothness parameter than for our marginal models.
The reason for this is that, while we expect temporal changes in the marginal distribution of precipitation to be very smooth,
we also expect the spatial distribution of extreme precipitation realisations to be highly localised and much less smooth.

Inference with \texttt{R-INLA} is performed within 1--4 hours for each conditioning site, using only
one core on the same Linux server as before. We evaluate the five model fits by estimating posterior
means of the same variables as in Subplot A of Figure~\ref{fig:empirical_conditional}, using
\(1000\) posterior samples of \(\bm \theta\). Subplot B displays these posterior means from the
model fit based on conditioning site nr.~2. Although there are some differences between Subplots A
and B, the patterns of the estimated curves are in general agreement, indicating a satisfactory
model fit overall.  The posterior mean of \(\mu(d; y_0)\) is similar to that of the data, with some
slight underestimation for large values of \(d\) and \(y_0\). The standard deviation
\(\zeta(d; y_0)\) is slightly underestimated for small \(d\), and overestimated for large \(d\). For
the values of \(\chi(d; y_0)\), which we care the most about, this results in a weak underestimation
for small \(d\) and large \(y_0\), and overestimation for large \(d\) and small \(y_0\).  We believe
that more complex models for \(a(\cdot)\) and \(b(\cdot)\), e.g., with \(\beta(\cdot)\) being a
function of \(y_0\) at small \(d\), would be able to further reduce the differences seen in
Figure~\ref{fig:empirical_conditional}. However, for the scope of this paper, we deem that the
current fit is good enough. We also believe that, when considering aggregated precipitation amounts inside a catchment area, the combination of minor overestimation and underestimation of the conditional exceedance probabilities at different distances might somewhat cancel each other out.
Estimates based on all
five model fits are displayed in Figure~\ref{fig:intensity_results_appendix} of the Supplementary Material, and they all seem to capture the major
trends of the data.

\subsection{Conditional occurrence process}%
\label{sec:case_study:occurrence}

\begin{figure}
  \centering
  \includegraphics[width=.99\linewidth]{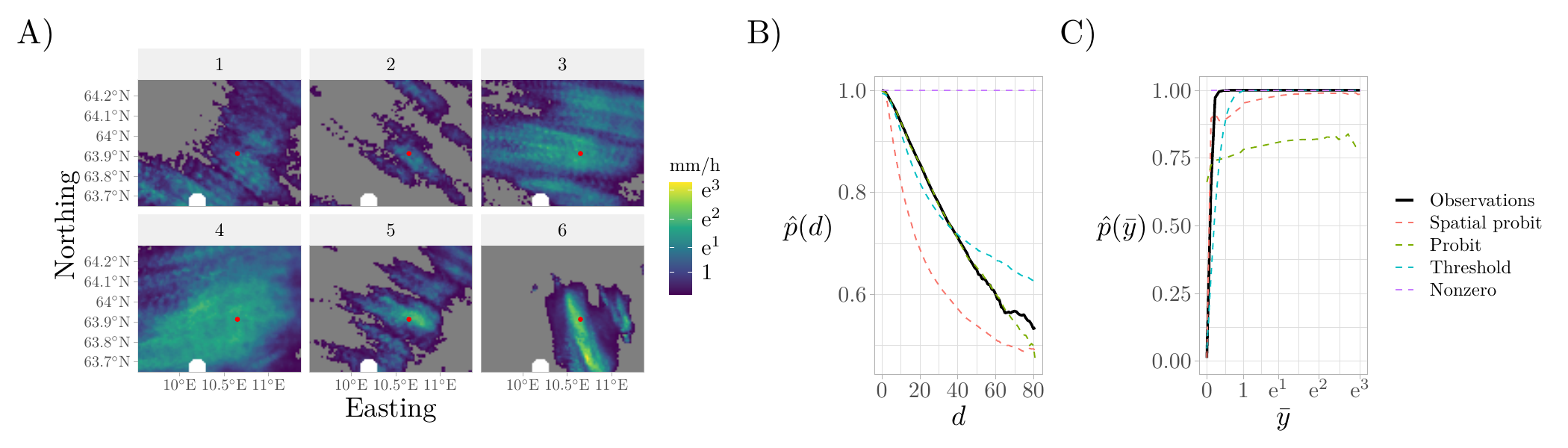}
  \caption{ A) Visualisation of observed precipitation for six time points where a threshold
    exceedance is observed at conditioning site nr.~4 (marked by a red dot). Grey denotes zero
    precipitation, while white denotes observations close to the Rissa radar that have been removed.
    The rightmost plots display empirical estimates for \(p(d)\) (B) and \(p(\bar y)\) (C), created
    using both observed data and simulated data from the four different occurrence models.}%
  \label{fig:zeros_exploratory}
\end{figure}

To model conditional precipitation occurrences, we first search for patterns in the observed
data. Since we model \([I_t(\bm s) \mid Y(\bm s_0) > \tau]\), we expect the occurrence probability
to be higher as we move closer to \(\bm s_0\). We therefore compute empirical occurrence
probabilities \(\hat p(d)\) at different distances \(d\) from \(\bm s_0\), using a sliding window of
width \(1\)~km. These are displayed by the black line in Subplot B of
Figure~\ref{fig:zeros_exploratory}. As expected, \(\hat p(d)\) decreases as \(d\) increases, with an
almost linear decline. Subplot A of Figure~\ref{fig:zeros_exploratory} displays six realisations of
the precipitation data. The distribution of precipitation occurrence appears to be smooth in space,
in the sense that zeros cluster together. Thus, the non-spatial probit model is unable to produce
realistic looking simulations. The precipitation intensities also appear to be smooth in space, in
the sense that we never observe big jumps in the precipitation values, and that zeros only occur
next to other zeros or small precipitation values. To check if this is true for all the available
data, we estimate the probability \(p(\bar y)\) of observing precipitation as a function of the mean
observed precipitation \(\bar y\) at the four closest spatial locations. The empirical estimates are
displayed using the black line in Subplot C of Figure~\ref{fig:zeros_exploratory}. It seems that the
probability of observing precipitation is close to zero if \(\bar y = 0\), and that it increases as
a function of \(\bar y\), and is almost exactly 1 if \(\bar y > 0.2\)~mm/h. This implies that our
probit models might produce unrealistic simulations, as they are independent of the intensity model
and might produce zeros close to large precipitation values.

Based on the exploratory analysis, we model the mean \(\mu(d)\), of the two probit models, using a
spline. More specifically, we model \(\partial \mu(d) / \partial d\) as a spline function based on
0-degree B-splines, where we place Gaussian zero-mean priors with a standard deviation of \(10\) on
all spline coefficients. Additionally, in the spatial probit model, we place PC priors on the SPDE
parameters \citep{fuglstad19_const_prior_that_penal_compl} such that the prior probability that the
range parameter \(\rho\) exceeds \(70\)~km is \(5\%\) and the prior probability that the standard
deviation of \(\mathcal W^*_t(\bm s; \bm s_0)\) exceeds \(5\) is \(1\%\). The smoothness parameter is once again fixed to
\(\nu = 0.5\). We then perform inference separately for each of the five chosen conditioning sites
and the two probit models. Inference with \texttt{R-INLA} takes between 10--15 minutes for the
spatial probit models, and 2--3 minutes for the non-spatial probit models. For the threshold model,
we estimate the threshold by computing the empirical probabilities of observing precipitation inside
the drainage area given extremes at each of the five conditioning sites.

We evaluate model performance by comparing properties of observed and simulated data. The threshold
model depends on the intensity process, so we first simulate conditional intensities, by sampling
\(\bm \theta\) from its posterior distribution and then sampling both
\([Y_t(\bm s_0) \mid Y_t(\bm s_0) > \tau]\) and
\(\{Y_t(\bm s) \mid Y_t(\bm s_0), \bm \theta: \bm s \in \mathcal S\}\)
using~\eqref{eq:intensity_model}. Then, we ``simulate'' occurrences with the threshold model by
rounding all small enough precipitation intensities down to zero. Figure~\ref{fig:zeros_exploratory}
displays empirical estimates for the probability of observing precipitation as a function of the
distance \(d\) to \(\bm s_0\) and as a function of the mean \(\bar y\) of the four nearest
neighbours, for observed and simulated precipitation data.  Clearly, the nonzero model fails to
capture the probability of precipitation occurrences. From Subplot B we find that the spatial probit
model heavily underestimates the probability of observing precipitation for most distances
\(d\). The threshold model performs better than the spatial probit model, but it slightly
underestimates \(p(d)\) for small \(d\) and overestimates it for large \(d\). The non-spatial probit
simulations, however, seem to agree well with the observed data for all values of \(d\). From
Subplot C of Figure~\ref{fig:zeros_exploratory}, we see that both probit models simulate zeros right
next to large precipitation observations, resulting in an underestimation of \(p(\bar y)\).
The classical probit model also often simulates precipitation occurences when little or no precipitation occurs at the neighbouring locations, thus overestimating \(p(\bar y)\) for small values of \(\bar y\). 
The
spatial probit model performs better than the classical independent one, but it still does not
completely solve this misfit.
The threshold occurrence simulations, however, seem to agree well with
the observations by placing its zeros close to other zeros or small precipitation values. Overall,
the threshold model seems to be the best at estimating occurrence probabilities. The classical
probit model is considerably better at estimating \(p(d)\), but it completely fails at estimating
\(p(\bar y)\).

\subsection{Simulating spatial precipitation extremes}%
\label{sec:case_study:final_simulations}

We combine all of the fitted models to simulate new realisations of extreme precipitation over the
drainage area.
For each of the five conditioning sites, we simulate \(N = {10}^3\)  scenarios of extreme precipitation over the same grid where radar data were observed. While our model is capable of simulating on a finer spatial scale, here we chose to use the same spatial resolution as the observed data to facilitate  comparison.

To create these scenarios,  we use 
Algorithm~\ref{alg:simulations},  where Exp\((1)\) denotes the
exponential distribution with unit scale, \(\mathcal S_1 \subset \mathbb R^2\) denotes all locations
where we simulate extreme precipitation, \(\hat F^{-1}_t\) denotes the estimated marginal quantile
function of positive hourly precipitation at time \(t\), and \(F\) denotes the cumulative
distribution function of the Laplace distribution. Recall that \(\mathcal T\) is the set of all time
points in our data. This algorithm can essentially be visualised by moving from the right to the
left in the model flowchart in Figure~\ref{fig:flowchart}.

\begin{figure}[h!]
  \centering
  \includegraphics[width=.99\linewidth]{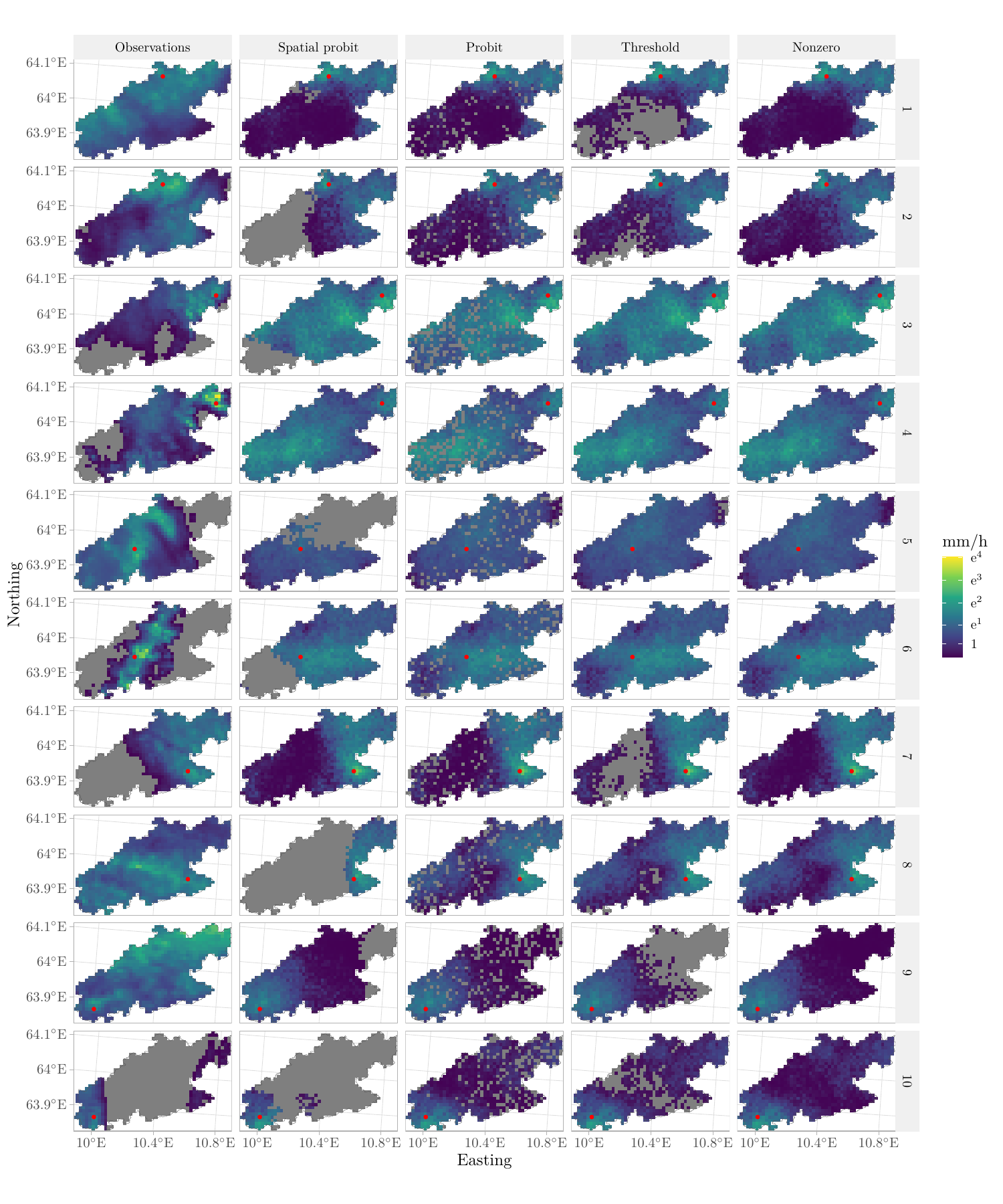}
  \caption{Realisations of conditional extreme precipitation, from observed and simulated data. The
    red dots display the locations of the chosen conditioning site for each subplot.}%
  \label{fig:precipitation_simulations}
\end{figure}

Figure~\ref{fig:precipitation_simulations} displays two observed and simulated realisations of
extreme precipitation over the drainage area for each of the five chosen conditioning sites and with
each of the four occurrence models. Simulations from the classical probit model and the nonzero
model are not capturing the spatial structure of precipitation occurrence in the observed data,
while simulations from the spatial probit model and the threshold model look more
realistic. However, unlike the threshold model simulations, many of the spatial probit simulations
contain large precipitation intensities right next to zeros, which is unrealistic.

\begin{algorithm}[t]
  \caption{Simulating spatial extreme precipitation with conditioning site \(\bm s_0\).}%
  \label{alg:simulations}
  \begin{algorithmic}
    \State Sample \(N\) time points \(t_1, \ldots, t_N\) uniformly with replacement from
    \(\mathcal T\).
    \For{\(i = 1, 2, \ldots, N\)}
    \State Sample a threshold exceedance \([Y_i(\bm s_0) \mid Y_i(\bm s_0) > \tau] \sim
    \tau + \text{Exp}(1)\)
    \State Sample a spatial realisation of the conditional intensity process \(\{Y_i(\bm s): \bm
    s \in \mathcal S_1\} \mid
    Y_i(\bm s_0)\)
    \State
    Sample a spatial realisation of the conditional occurrence process \(\{I_i(\bm s): \bm s \in
    \mathcal S_1\} \mid Y_i(\bm s_0)\)
    \State 
    Transform back to the precipitation scale: \(X^+_i(\bm s) =
    \hat F^{-1}_{t_i}[F(Y_i(\bm s))]\)
    \State Add zeros to the samples: \(X_i(\bm s) = X^+_i(\bm s) I_i(\bm s)\)
    \EndFor
  \end{algorithmic}
\end{algorithm}

\begin{figure}
  \centering
  \includegraphics[width=.99\linewidth,page=1]{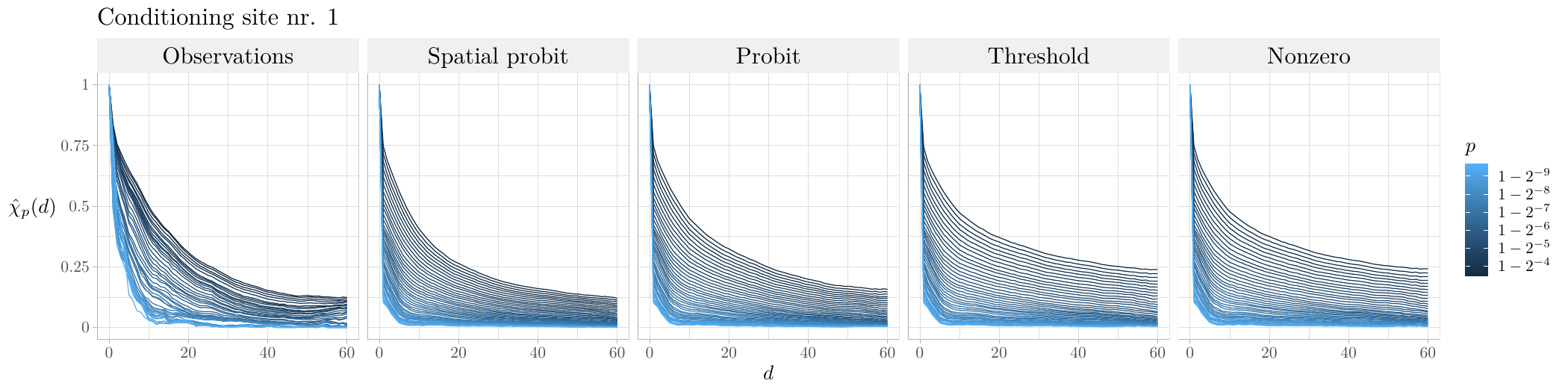}
  \caption{Estimates of \(\chi_p(d)\) for observed and simulated data, using conditioning site
    nr.~1.}%
  \label{fig:precipitation_chi}
\end{figure}

As discussed in Section~\ref{sec:introduction}, both the amount of precipitation and its spatial
distribution are important features for assessing flood risk. Thus, to further evaluate the
simulations, we compare conditional exceedance probabilities and precipitation sums over different
areas between the observed and simulated data. Estimates of \(\hat \chi_p(d)\) are computed using
the same sliding window approach as in Section~\ref{sec:case_study:intensity}.
Figure~\ref{fig:precipitation_chi} displays the estimates from conditioning site nr.~1. The
simulations seem to capture \(\chi_p(d)\) well, even though, just as in
Section~\ref{sec:case_study:intensity}, \(\chi_p(d)\) is somewhat underestimated for small
distances, \(d\), and overestimated for large distances. The probit models seem to overestimate
\(\chi_p(d)\) less than the non-probit models at large distances. This makes sense because the
probit models are independent of the intensity process and may therefore set large intensity values
equal to zero. The estimates of \(\chi_p(d)\) seem almost identical for the threshold model
simulations and the nonzero model simulations.  This also makes sense, since the small values that
are rounded down to zero by the threshold model are too small to considerably affect the values of
\(\chi_p(d)\). Similar patterns are found for the other four conditioning sites. Estimates of
\(\chi_p(d)\) from all five conditioning sites are displayed in Figure~\ref{fig:precipitation_chi_supplement} of the Supplementary Material. 

\begin{figure}
  \centering
  \includegraphics[width=.99\linewidth,page=4]{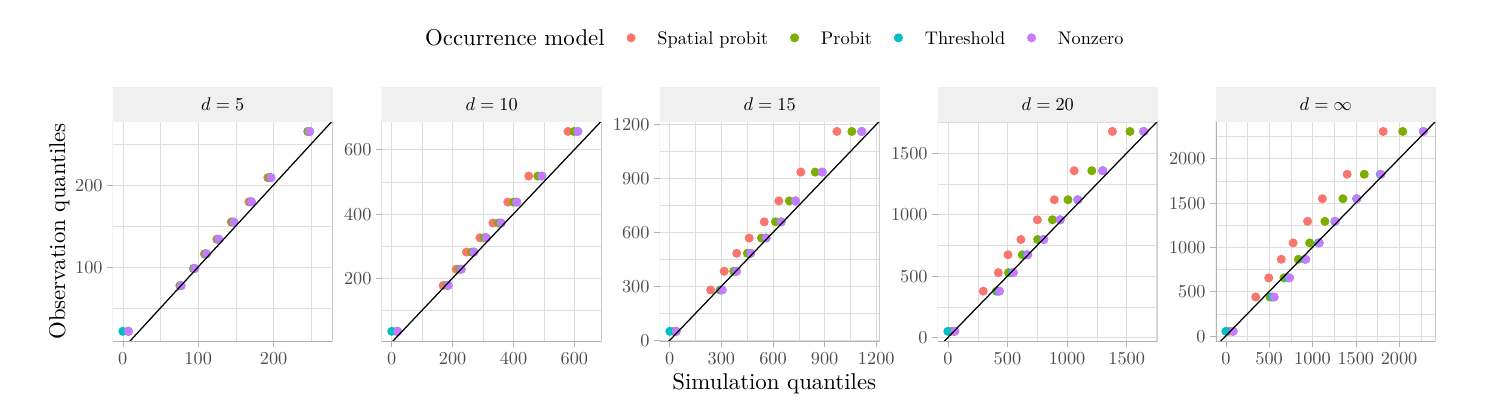}
  \caption{QQ plots for the sum of aggregated precipitation inside
    \(\mathcal B_d(\bm s_0) \cap \mathcal S_1\), based on conditioning site nr.~4. Quantiles from
    the four different simulated data sets are displayed using different colours.}%
  \label{fig:qq_sums}
\end{figure}

We also compare aggregated simulated and observed precipitation amounts over the drainage area to
evaluate the simulations. For each conditioning site, we compute precipitation sums inside
\(\mathcal B_d(\bm s_0) \cap \mathcal S_1\), where \(\mathcal B_d(\bm s_0)\) is a ball of radius
\(d\)~km, centred at \(\bm s_0\), and \(\mathcal S_1\) denotes the catchment of interest. We then
compare observed and simulated precipitation sums using QQ plots. Figure~\ref{fig:qq_sums} display
these plots for conditioning site nr.~4. For small \(d\), all the simulations produce similar
precipitation amounts, which are close to the observed data, although slightly smaller. As \(d\)
increases, the underestimation increases somewhat for the probit models, while it decreases for the
non-probit models, which seem to agree well with the observed data. Quantiles of the threshold model
and the nonzero model are almost identical and can be hard to distinguish. QQ plots for all five
conditioning sites are displayed in Figure~\ref{fig:qq_sums_supplement} of the Supplementary Material. They all display similar patterns.

\section{Discussion and conclusion}%
\label{sec:conclusion}

We propose a framework for modelling and simulating high-dimensional spatial precipitation extremes,
which focuses on describing the marginal distributions, spatial tail dependence structure and
intermittency of extreme high-resolution precipitation data. We consider this as a first step in the
development of improved stochastic weather generators, which can be of great use for, e.g.,
hydrological impact assessment studies over small or urban catchments or as input scenarios for so-called serious games. 

Our framework is built using the spatial conditional extremes model and latent Gaussian models. We
model the marginal distributions of nonzero precipitation using a mixture of two latent Gaussian
models with a gamma likelihood and a generalised Pareto (GP) likelihood, while separately modelling
the extremal dependence of nonzero precipitation extremes with a spatial conditional extremes model
formulated as a latent Gaussian model. Precipitation occurrences are modelled using four different
competing binary models. Fast approximate inference is achieved using integrated nested Laplace
approximations (INLA). We develop new empirical diagnostics for the threshold and standardising
functions of the spatial conditional extremes model, and we use these to propose a new class of
parametric forms for the standardising functions. The framework is applied for simulating
high-resolution precipitation extremes over a water catchment in Central Norway, and is found to
efficiently produce spatially reliable realisations of extreme precipitation.

The required choice of a conditioning site when using the conditional extremes model is somewhat
controversial in the field of extreme value analysis. Some view it as a deficiency, as it  makes it
impossible to build a global model without having to condition on ``arbitrary conditioning sites''. 
We, instead,  view it more as a strength, as it lets us examine how properties of the extreme field
change as the ``center of the storm'' moves around inside our spatial domain of interest. Moreover,
it is the use of a conditioning site that allows the spatial conditional extremes model to provide
computationally efficient high-dimensional descriptions of subasymptotic dependence structures, that
are flexible enough to provide reliable models for complex phenomena like extreme spatial
precipitation fields. Without this conditioning, we would not, to the best of our knowledge, have
been able to provide as flexible and computationally efficient descriptions of high-resolution
extreme precipitation fields.

The threshold occurrence model appears to outperform the other occurrence models. It captures
both spatial and marginal properties of the data well, while the two probit models fail to capture
either the marginal or the spatial properties. However, since the precipitation simulations stem
from a combination of intensity samples and occurrence samples, it is nontrivial to conclude that
one occurrence model significantly outperforms the others, as a change in the intensity model might
potentially cause another occurrence model to produce better precipitation simulations.

Compared to the probit models, the threshold model lacks some flexibility in the sense that it
produces deterministic simulations given the intensity samples and a threshold. However, the
threshold model interacts with the intensity model, which the probit models are unable to. This
interaction is clearly crucial, as the threshold model ends up being our most successful. For future
work, it might prove fruitful to develop probit models that interact more with the intensity
model. A common approach for creating such interactions is to perform joint inference, where the two
models share some latent model components, similar to \citet{bacro20_hierar_space_time_model_asymp}
and \citet{gelfand21_multiv_spatial_proces_model}. This might be challenging for the intensity and
occurrence models, as their latent fields have different interpretations and scales, but it might
still be possible to create some meaningful link between the two.

The spatial probit model underestimates occurrence probabilities almost everywhere in space. We
believe this happens because the spatial clustering of zeros and ones in the data forces the nugget
effect to be small, which, in the latent Gaussian model formulation, causes the latent field to have
a large enough variance to absorb most of the mean trend. The symmetry of the latent field thus
makes all marginal probabilities tend towards \(50\%\), meaning that it underestimates large
probabilities and overestimates small probabilities. One might fix this by removing the conditional
independence assumption, i.e., discarding the nugget effect. However, this makes inference with INLA
impossible. Alternatively, one might use an asymmetric latent random field, with a skewness that
varies in space, so that latent variables close to the conditioning site are right skewed, while
latent variables further away from the conditioning site are more left
skewed. \citet{cabral24_fittin_laten_non_gauss_model} show that inference with INLA can be possible for
non-Gaussian latent fields, meaning that \texttt{R-INLA} might work with such a model.
Additionally, since the standardised intensity process has Laplace marginals, the conditional
intensity model should not impose Gaussian marginals on observations that are independent of the
conditioning site. \citet{wadsworth22_higher_spatial_extrem_singl_condit} solve this problem by
modelling the residual field of the spatial conditional extremes model using delta-Laplace marginal
distributions, which encompass both Gaussian and Laplace marginal distributions. The work of
\citet{cabral24_fittin_laten_non_gauss_model} might make it possible to model the conditional
intensity
process using a non-Gaussian latent field with delta-Laplace marginal distribution in INLA.

As discussed in Section~\ref{sec:introduction}, radar data are an excellent source of information for capturing the small-scale spatio-temporal variability of precipitation. In fact, they may be virtually the only data available that can provide such detailed information. However, radar data have not been widely used for modelling extreme precipitation. We show that the radar data let us capture small-scale extremal dependence structures
with high precision. This is a promising result, as weather radar data are easily available over
several basins where rain gauges might be very sparse or absent. Despite this, weather radars are known
to struggle with accurately capturing the exact precipitation amounts, i.e., marginal distributions. Transforming reflectivity into precipitation amounts can introduce  artefacts, which may explain why our marginal models did not detect any considerable elevation trends in the data. This
might negatively affect our estimates of aggregated precipitation amounts. To generate reliable
simulations of extreme precipitation, one should attempt to combine information from multiple
precipitation data sets using data-fusion approaches. For example, one could model extremal dependence using high-density radar data,
while using rain gauge observations to model marginal distributions, as they better  capture
precipitation amounts, despite being too sparse in space to provide successful
estimates of the extremal dependence structure. To properly evaluate the model performance, one
should also evaluate model performance by using the precipitation simulations as input to
hydrological models, and by examining their effects on design flood estimation methods.

Threshold selection for the GP distribution is a complicated problem \citep[e.g.,][]{coles01_introd_statis_model_extrem_values, Scarrott_MacDonald_2012, ROTH20141,  silva20_l_momen_autom_thres_selec},
especially when considering high-dimensional and/or spatio-temporal data sets.
To the best of our knowledge, there is no general agreement in the literature about how to best
select or evaluate a threshold for the GP distribution in such complex settings,
and the most common method is therefore to choose a threshold that is believed to provide a good trade-off between the model variance and model calibration for both extreme and sub-extreme 
data~\citep[e.g.,][]{youngman16_geost_extrem_value_framew_fast, castro-camilo19_splic_gamma_gener_paret_model, SimpsonEtAl2023Highdimensionalmodeling}.
We follow this approach and set the GP threshold equal to the \(95 \%\) quantile of the fitted marginal distributions,
which we believe to provide a suitable trade-off between a good model fit for the rightmost distributional tail and the model variance. 
However, we did not perform a thorough evaluation of our threshold choice.
\citet{opitz18_inla_goes_extrem} choose their threshold using cross-validation, but this would be computationally intractable for our modelling framework with the current data set. 
It is also not clear how changes in the marginal models, such as the GP threshold, might affect the performance of the conditional intensity and occurrence models.
As an example, it might be that one specification of the intensity model leads to \(p_u = 0.95\) performing the best in a cross-validation study, but that a small change in the intensity model leads to \(p_u = 0.99\) being the best threshold choice.
It would therefore be of interest to further examine how to best choose and evaluate the threshold for the GP distribution, and on how changes in the marginal models might affect the performance of the intensity and occurrence models.
Our framework could potentially be improved by exchanging the threshold-dependent GP distribution with a threshold-independent distribution such as the extended GP distribution of \citet{papastathopoulos13_exten_gener_paret_model_tail_estim} and \citet{naveau16}.

Our models assume isotropy, but the observed data in Figures~\ref{fig:zeros_exploratory}
and~\ref{fig:precipitation_simulations} display some indications of anisotropy. This does not seem
to affect our results much, as the simulated data capture the main trends of the observed data
well. However, in regions where topography strongly influences the storm direction patterns or there
are dominating storm directions (for example, low pressure along the Norwegian west coast usually
arrives from the south west), anisotropy can be more pronounced. An interesting avenue for future
research is therefore the addition of anisotropy and/or nonstationarity into the modelling framework.
\citet{wadsworth22_higher_spatial_extrem_singl_condit} account for anisotropy by transforming the
coordinates of their data before performing inference with the spatial conditional
extremes. \citet{lindgren23_diffus_based_spatio_tempor_exten} also demonstrate how to account for
anisotropy directly inside the SPDE model, which works well with the \texttt{R-INLA} framework.

We model the location function \(a(\cdot)\) of the spatial conditional extremes model using 
a fully parametric function. This is different from the approach of \citet{SimpsonEtAl2023Highdimensionalmodeling},
who model \(a(\cdot)\) using a semi-parametric one-dimensional Gaussian random field.
In our early attempts at modelling the intensity process we applied a similar semi-parametric \(a(\cdot)\)-function.
However, we struggled with extracting reliable inferences using this model, and therefore switched to the fully parametric model, which we found to be more robust at the cost of being less flexible and less computationally efficient.
Since this change we have made several model improvements, including the constraining method of \citet{VandeskogEtAl2024efficientworkflowmodelling}, the specification of the \(b(\cdot)\)-function, and improvements to the marginal standardisation models. 
All these changes have led to more robust and efficient inferences.
With the added model improvements it might be possible to employ the semi-parametric \(a(\cdot)\)-function of \citet{SimpsonEtAl2023Highdimensionalmodeling} for a more flexible modelling approach, 
without having an overly negative effect on the model robustness.
To the best of our knowledge, a similar semi-parametric model for \(b(\cdot)\) is not possible to achieve within the \texttt{R-INLA} framework.

It is known that \texttt{R-INLA} can struggle to approximate the posterior distribution if given
suboptimal initial values, or if some parameters are not well identifiable in practice. Our chosen
model for the conditional intensity process is highly flexible, and different combinations of the
model parameters may sometimes produce similar likelihood values. In practice, we have seen that
small changes in the model formulation or initial values can lead to large changes in the estimated
parameters, and care should therefore be taken when applying this methodology in other
settings. However, since these different parameters produce similar likelihood values, they all seem
to perform equally well, when considering the QQ plots and estimates of \(\chi_p(d)\) in
Section~\ref{sec:case_study:final_simulations}. We have never observed that a small change in model
formulation or initial values can lead to a noticeably worse model fit overall.

Parameters of the marginal precipitation distributions are estimated using latent Gaussian models
with conditional independence assumptions given a relatively simple latent field. Such assumptions
might fail to account for the complex spatio-temporal dependence structures of precipitation data
and might therefore produce too small uncertainty estimates, due to an overestimation of the
effective sample size.
Furthermore, no uncertainty propagation takes place between the different models.
However, to the best of our knowledge, no computationally tractable methods
exist that can account well for uncertainty propagations and such complex spatio-temporal dependence in such large and
high-dimensional data sets. Additionally, an underestimation of the uncertainty is not too
problematic when we only use point estimates of the parameters for modelling the marginal
distributions. Also, the reasonable parameter estimates displayed in
Section~\ref{sec:case_study:marginal_model} and the model evaluation in the Supplementary
Material imply that our marginal models perform well, even though they are based on the oversimplified
conditional independence assumption. We also assume the tail parameter for the GP distribution to be positive. This is appropriate for modelling precipitation, which is known to have a heavy tail.
Currently, \texttt{R-INLA} cannot fit GP models with a negative tail parameter.
Therefore, if the data suggest a possibly negative tail parameter, a different inferential method should be used for the GP model.

The conditional intensity and occurrence models are purely spatial, and they assume that
observations from different time points are independent, which can lead to too small uncertainty
estimates.
Thus, the next step towards an improved stochastic weather generator for extreme precipitation
should include temporal modelling components, as this might allow for better uncertainty
quantification, and might also make it possible to simulate the evolution and duration of an extreme
precipitation event. We do not believe that the inclusion of temporal components will entail too
much work, as extensions from space to space-time can be relatively simple to achieve within the spatial
conditional extremes model and the \texttt{R-INLA} framework. As an example,
\citet{SimpsonEtAl2023Highdimensionalmodeling} successfully perform both spatial and spatio-temporal
modelling with the spatial conditional extremes model, and it should be possible to extend most of
their changes for space-time modelling into our developed framework.

\section*{Acknowledgements}

\noindent \textbf{Funding} Rapha{\"e}l Huser was partially supported by funding from King Abdullah University of Science and
  Technology (KAUST) Office of Sponsored Research (OSR) under Award No.~OSR-CRG2020-4394

\noindent \textbf{Conflict of interest} The authors report there are no competing interests to declare.

\printbibliography%

\setcounter{section}{0}
\renewcommand{\thesection}{S\arabic{section}}
\numberwithin{equation}{section} 
\numberwithin{figure}{section} 
\numberwithin{table}{section} 

\section*{\huge Supplementary material}

\section{Data exploration}
\label{sec:supp_data_exploration}

\begin{figure}
  \centering
  \includegraphics[width=.99\linewidth]{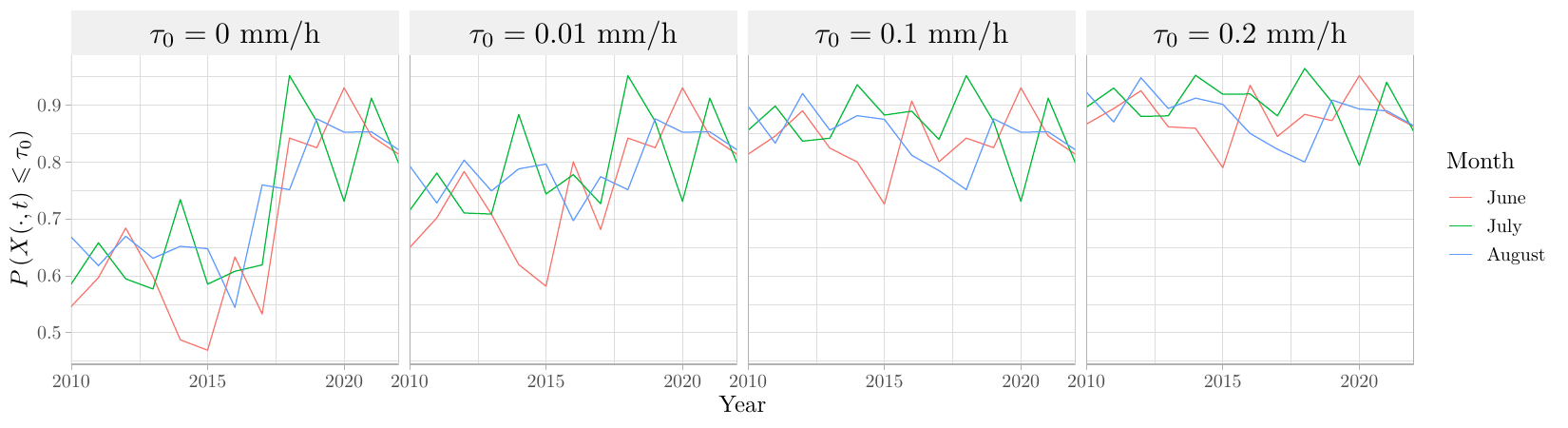}
  \caption{
    The temporal distribution of the proportion of observations, pooled across space, that are less than
    or equal to \(\tau_0\) for June, July and August, respectively.}%
  \label{fig:small_proportions}
\end{figure}

We compute the proportions of exact zeros in the precipitation data, pooled across space, for
different times. Figure~\ref{fig:small_proportions} displays temporal distributions of the
proportions of observations that are less than or equal to a threshold \(\tau_0\), for different
values of \(\tau_0\). The Rissa radar was upgraded in 2018 (personal communication, 2023), and this
is clearly visible from the lower left subplot, as the proportion of exact zeros changed
considerably in that year. In order to remove these changing zero-patterns, we post-process the
data by rounding every observation smaller than \(0.1\)~mm/h down to zero, as this seems to give a
somewhat equal proportion of zeros everywhere in space and time.

\section{Evaluation of fitted marginal distributions}
\label{sec:supp_evaluation}

\begin{figure}
  \centering
  \includegraphics[width=.8\linewidth]{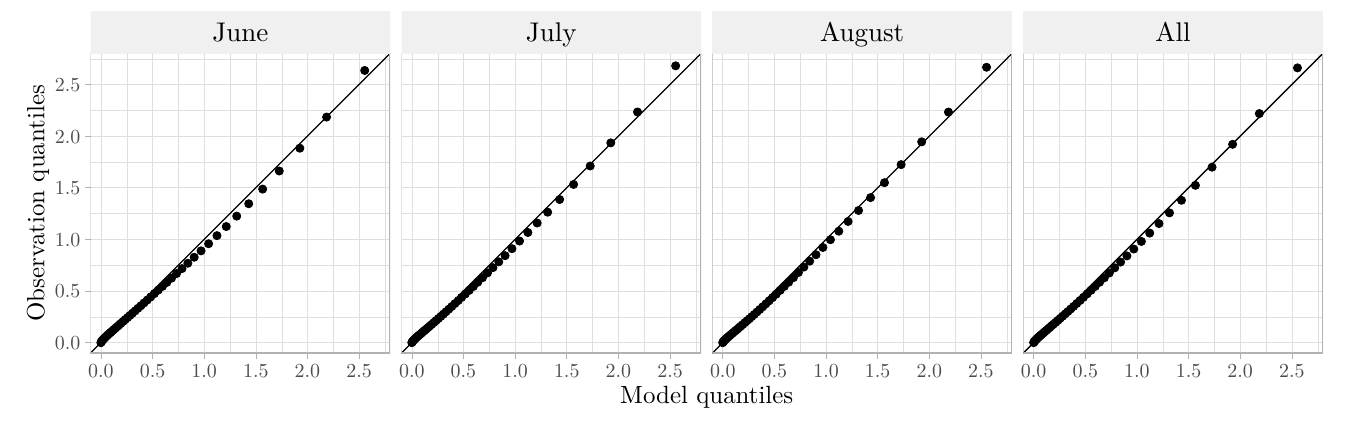}
  \includegraphics[width=.8\linewidth]{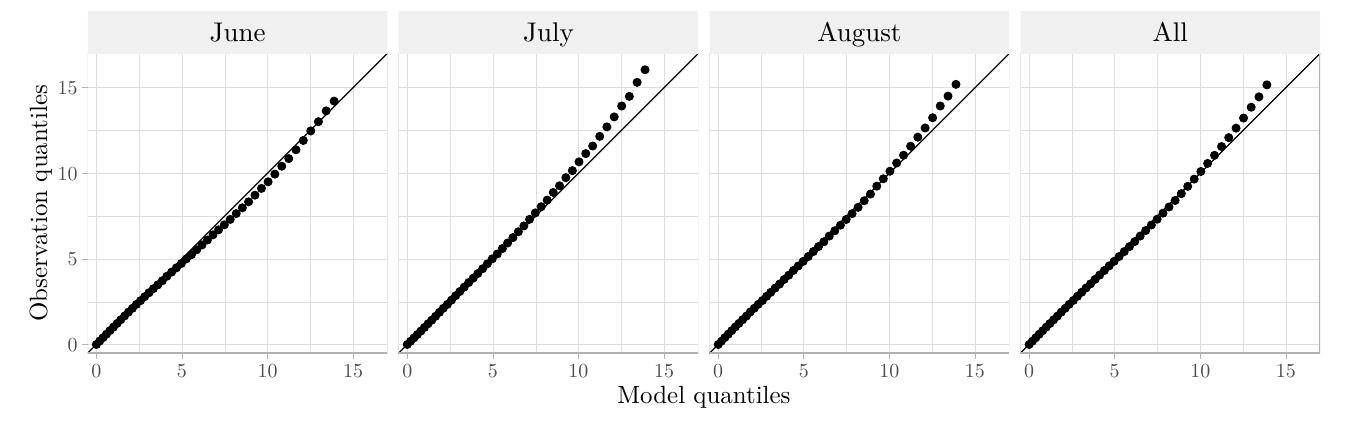}
  \caption{Upper row: QQ plots comparing empirical quantiles of the standardised observations with
    quantiles of a gamma distribution with a scale of 1 and the estimated shape parameter.
    Lower row: QQ plots comparing empirical quantiles of the standardised threshold exceedance
    observations with quantiles of a GP distribution with a scale of 1 and the estimated tail
    parameter.
    The four columns display empirical quantiles for June, July, August and for all the
    months together, respectively.}%
  \label{fig:qq}
\end{figure}

We evaluate goodness of fit for the two marginal models by creating quantile-quantile (QQ) plots.
It is not straightforward to compare empirical quantiles with our model quantiles, as the models
assume that each day comes with a different distribution, and thus a different set of quantiles.  To
create QQ plots for the gamma model, we therefore standardise the data by dividing each observation
by its estimated scale parameter. Then we compare empirical quantiles of the standardised data with
quantiles from a gamma distribution with a scale parameter of 1 and the estimated shape
parameter. The resulting QQ plots are displayed in the upper row of Figure~\ref{fig:qq}.

The QQ plots do not contain any uncertainty estimates.
The reason for this is that the conditional independence assumption of the two latent Gaussian models can lead to 
a considerable underestimation of model variance if the latent fields fail to capture the entire spatio-temporal dependence structure. 
The uncertainties of the estimated model quantiles are therefore unrealistically small and would provide little information about the actual model uncertainties.
Furthermore, as the precipitation data are both spatially and temporally correlated, it is not trivial to estimate the uncertainties of the empirical quantiles of the standardised data either.
Note that the underestimation of the marginal model variance is not a problem for us when performing the actual modelling in the main paper,
because we only use the estimated posterior means of the model parameters when performing the marginal standardisation.
These are unaffected by the underestimation of the model variance~\citep[e.g.][]{ribatet12_bayes_infer_from_compos_likel, VandeskogEtAl2024efficientworkflowmodelling}.

Similarly to the QQ plots for the gamma model,
in order to create QQ plots for the generalised Pareto (GP) model, we standardise the observed threshold
exceedances by dividing them by their estimated scale parameters, and we then compare empirical quantiles
with those of a GP distribution with a scale of 1 and the estimated tail parameter. The QQ plots are
displayed in the lower row of Figure~\ref{fig:qq}.

The GP model slightly underestimates the largest quantiles, but we here note that a value of
10~mm/h is so large that it corresponds to the empirical \(99.3\%\) quantile of the standardised
threshold exceedances, which is approximately the same as the \(99.97\%\) quantile of the nonzero
precipitation observations, and approximately the same as the \(1 - 5 \times 10^{-5}\) quantile of
the hourly summer precipitation observations. This is slightly more than the 9 year return
level for summer precipitation under the (unlikely) assumption that all observations are i.i.d.

It is difficult to say with certainty what causes this underestimation.
It might be that the underestimation can be reduced by changing the gamma model through, for example,
choosing a more flexible linear predictor,
using a different likelihood,
censoring observations above the GP threshold within the gamma likelihood,
propagating uncertainty by using more information than the posterior mean of the model parameters,
or by removing the conditional independence assumption of the model.
It might also be possible to reduce underestimation by making changes to the GP model,
such as choosing another GP threshold than the \(95\%\) marginal quantile,
using a more flexible model for the linear predictor,
or by removing the conditional independence assumption,
which can be especially problematic if the extremes in the data tend to cluster~\citep[e.g.,][]{ROTH20141, davison19_spatial_extrem}.
We attempt to model the marginal distributions using the two other GP thresholds \(p_u = 0.90\) and \(p_u = 0.98\). 
Apart from a minor improvement in the August quantiles for the \(p_u = 0.90\) threshold, no considerable differences are seen in the resulting QQ plots for the other two marginal models.
The changes in the estimated tail parameter \(\xi\) between the three fitted marginal models are also small.
All three estimates are between \(0.14\) and \(0.15\), and the differences between the estimates are on an order of magnitude \(\sim 10^{-3}\).
It should also be noted that the large uncertainty of high upper order statistics, combined with the uncertainty of the estimated model quantiles, might make the underestimation appear more considerable than it actually is.
It would be interesting in future research to better understand whether the underestimation of the largest quantiles can be removed by implementing some or all of the proposed improvements to the marginal model,
and how much of the underestimation that is simply created by random chance,
due to the hidden uncertainty of the largest quantiles in the QQ plots.

\begin{figure}
  \centering
  \includegraphics[width=.9\linewidth]{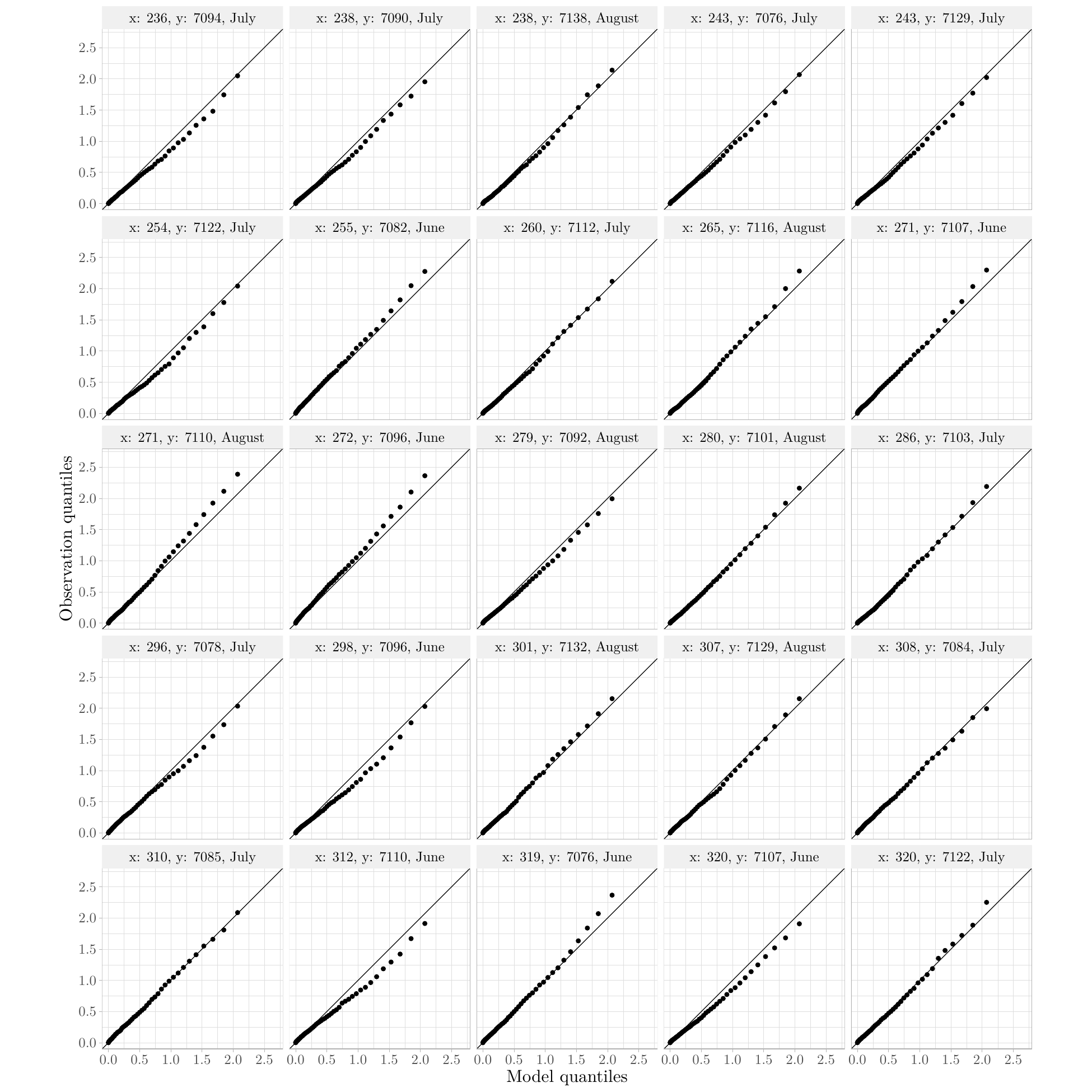}
  \caption{QQ plots comparing empirical quantiles of the standardised observations with
    quantiles of a gamma distribution with a scale of 1 and the estimated shape parameter, for 25 randomly sampled combinations of spatial location and month.}%
  \label{fig:qq_gamma_random}
\end{figure}

\begin{figure}
  \centering
  \includegraphics[width=.9\linewidth]{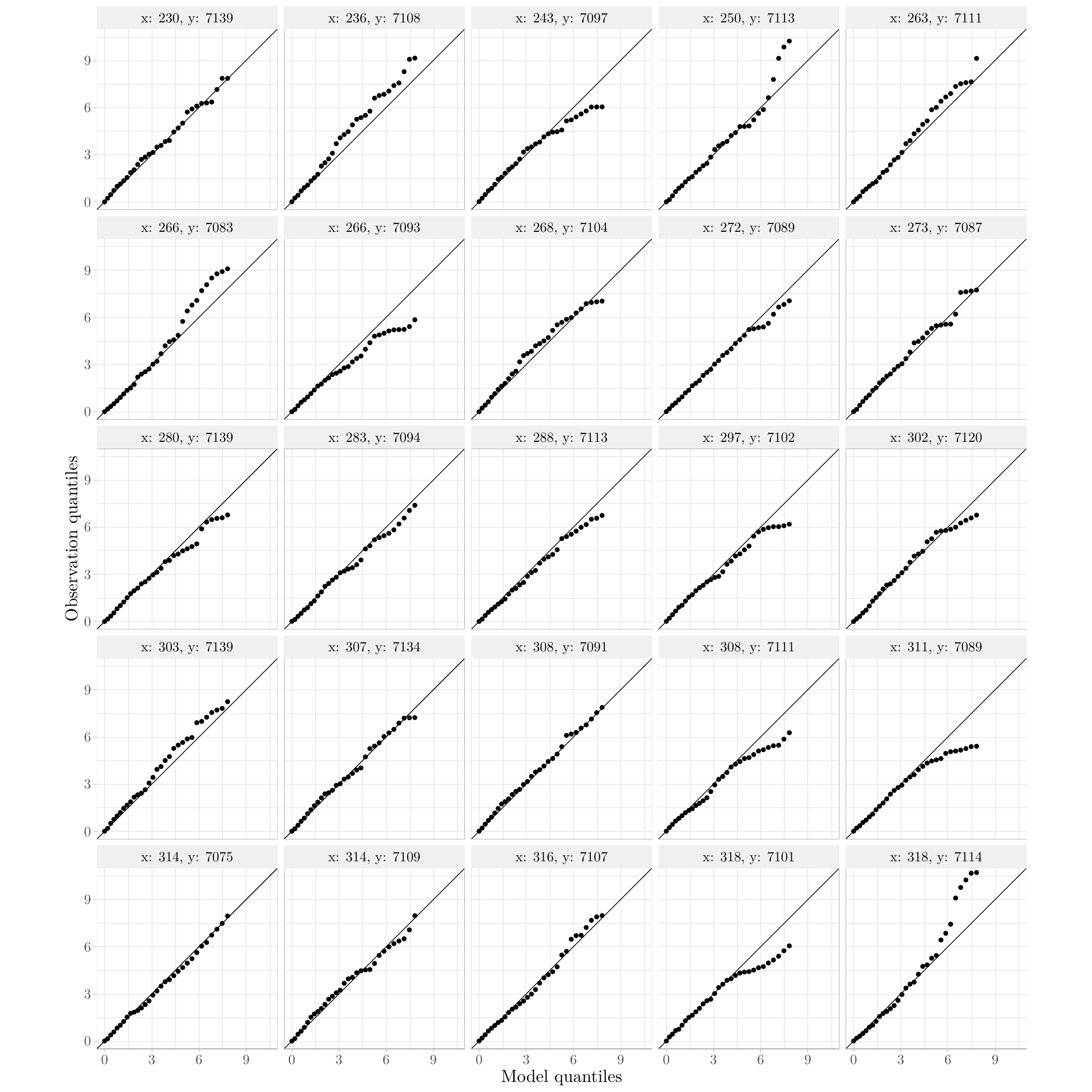}
  \caption{QQ plots comparing empirical quantiles of the standardised threshold exceedance
    observations with quantiles of a GP distribution with a scale of 1 and the estimated tail
    parameter, for 25 randomly sampled combinations of spatial location.}%
  \label{fig:qq_gp_random}
\end{figure}

Highly aggregated model evaluations, such as the QQ plots in Figure~\ref{fig:qq}, might hide more localised model deficiencies in time or space, as
a model that is underdispersed in half of the spatial locations and overdispersed in the other half might still produce a good-looking QQ plot.
To further evaluate the more localised performance of our model, we 
also examine QQ plots for smaller subsets of the data.
Figure~\ref{fig:qq_gamma_random} displays QQ plots for 25 randomly sampled combinations of spatial location and month.
Some of the subplots display model deviations that are, as expected, larger than in the upper row of Figure~\ref{fig:qq}. However, the QQ plots still demonstrate a good model fit at most locations and months.
Similar patterns have been found for all other location-month combinations that we have examined. 
The procedure is repeated for the fitted generalised Pareto model.
However, here we choose to pool together data from all three summer months, as the total number of threshold exceedances per location,
over the entire time period, only ranges from 180 to 280, with a mean of 227.
Figure~\ref{fig:qq_gp_random} displays QQ plots for 25 randomly sampled spatial locations. 
Once more, the model deviations are larger than in the pooled QQ plots in Figure~\ref{fig:qq}, but the plots imply a quite satisfactory model fit overall.
Similar patterns have been found for all other spatial locations that we have examined.

\begin{figure}
  \centering
  \includegraphics[width=.9\linewidth]{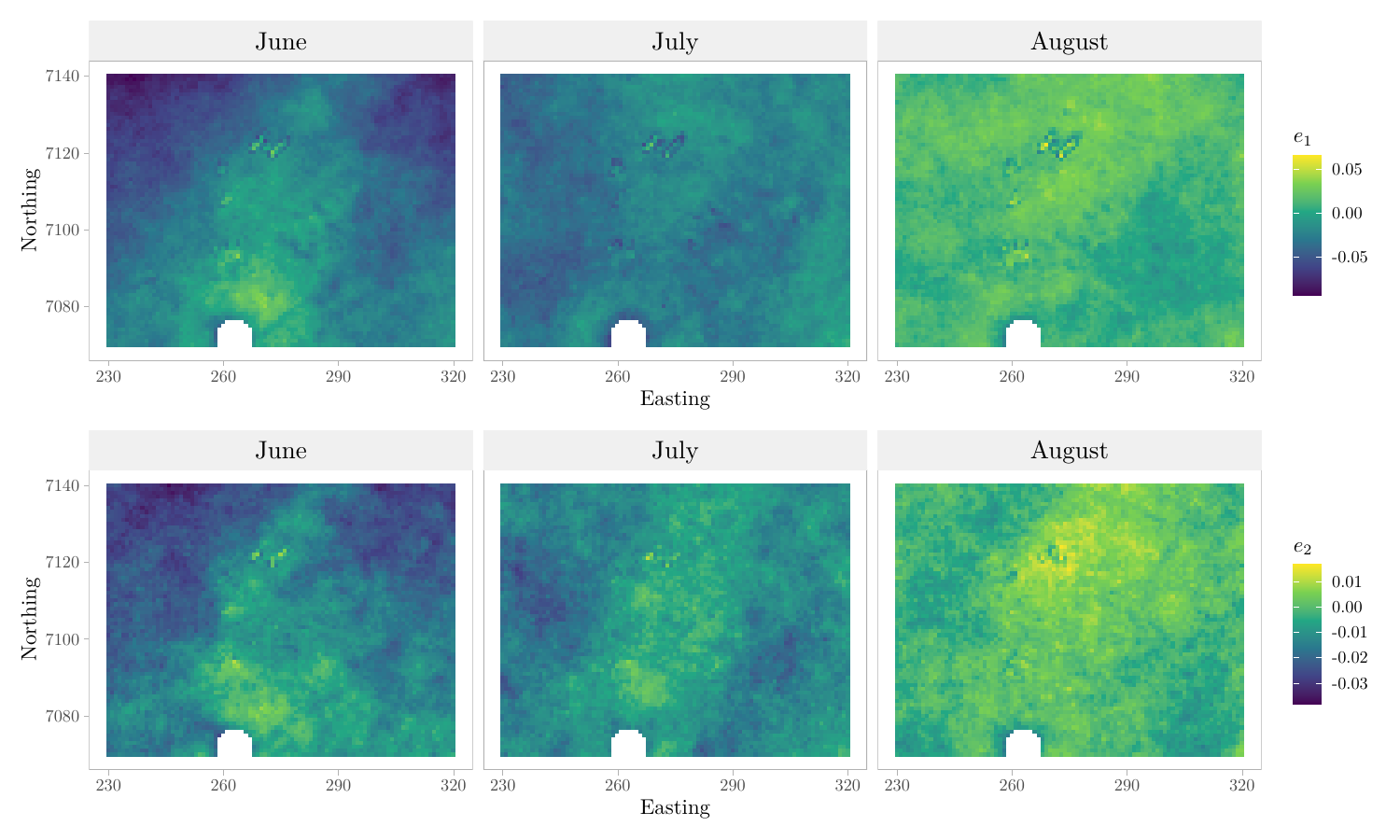}
  \caption{PIT errors \(e_1\) and \(e_2\), for the fitted marginal model, computed for each location/month combination of the available data. The white area at the bottom of the spatial domain contains all data that were removed due to distortions close to the Rissa radar, as described in the data section of the main paper.
  }%
  \label{fig:pit}
\end{figure}

\begin{figure}
  \centering
  \includegraphics[width=.8\linewidth]{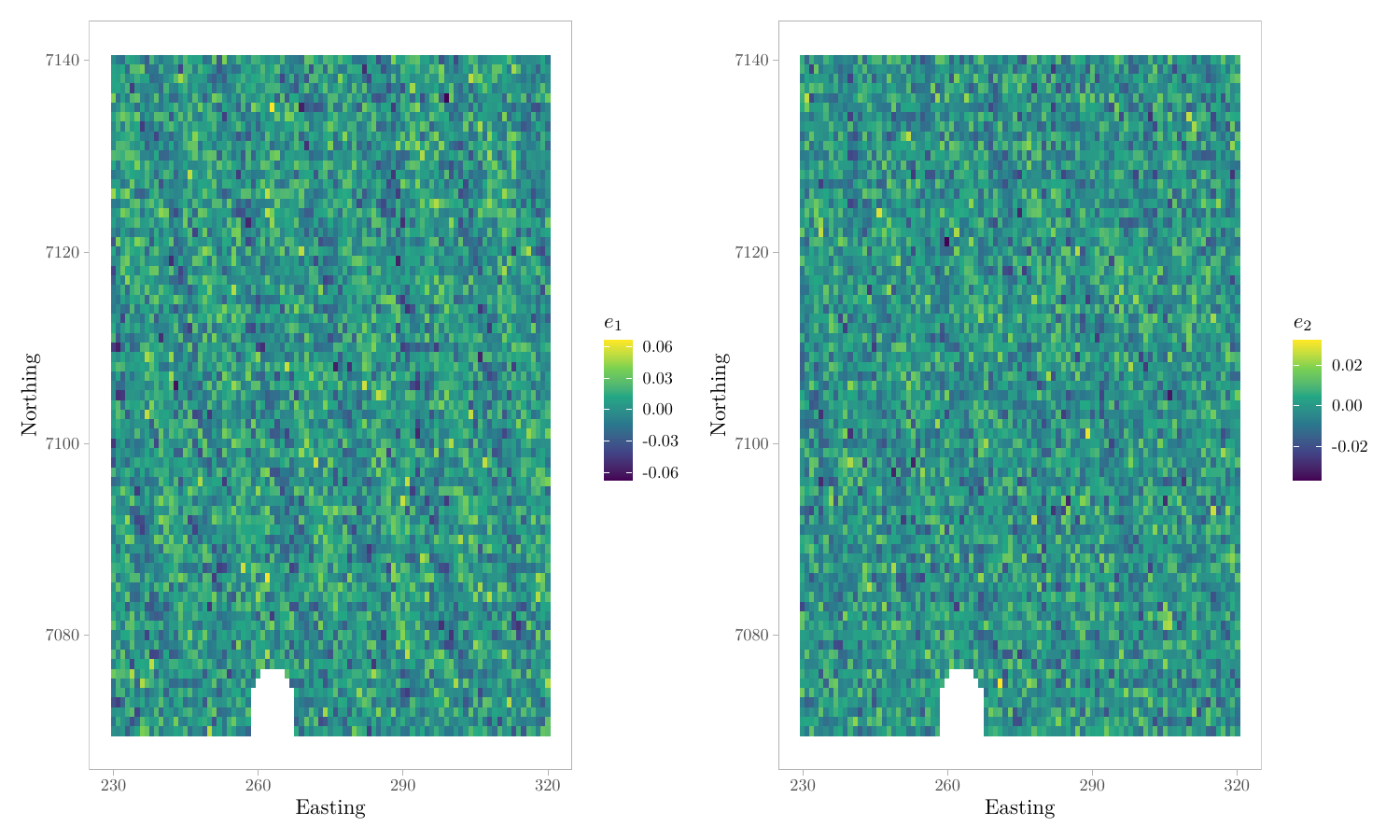}
  \caption{PIT errors \(e_1\) and \(e_2\), for the fitted generalised Pareto tails of the marginal model, computed for each location/month combination of the available data. The white area at the bottom of the spatial domain contains all data that were removed due to distortions close to the Rissa radar, as described in the Section~\ref{sec:data} of the main paper.
  }%
  \label{fig:gp_pit}
\end{figure}

Even further model evaluation is performed by using our fitted marginal distributions
to transform all the data to be approximately uniformly distributed between 0 and 1, using the probability integral transform (PIT).
Then, similarly to \citet{HeinrichEtAl2020MultivariatePostprocessingMethods}, we evaluate model calibration by comparing empirical moments of the data to empirical moments of the uniform distribution.
If the model is well calibrated, the transformed data should have an empirical mean close to \(0.5\) and an empirical standard deviation close to \(1 / \sqrt{12} \approx 0.29\). 
Too small empirical means indicate positive model bias, while too large empirical means indicate negative model bias.
Similarly, too small empirical standard deviations indicate an overdispersed model, while too large empirical standard deviations indicate an underdispersed model.
We therefore compute the PIT errors \(e_1 = m - 0.5\) and \(e_2 = s - 1 / \sqrt{12}\), where the empirical mean and standard deviation of the transformed data are denoted by \(m\) and \(s\) respectively.
The PIT errors \(e_1\) and \(e_2\) are displayed in Figure~\ref{fig:pit}.
The model seems to be sligthly positively biased and overdispersed during June, and slightly negatively biased and underdispersed during August.
There also appear to be some slight spatial trends in the PIT errors.
However, both the spatial and the temporal trends are relatively mild, and
the absolute value of \(e_1\) is smaller than \(0.045\) for \(90\%\) of the month/location combinations, while
the absolute value of \(e_2\) is smaller than \(0.021\) for \(90\%\) of the month/location combinations.
This implies that there are no areas in space-time where considerable discrepancies are present.
A similar analysis is performed for the data that are large enough to be modelled by the generalised Pareto distribution.
Just as for the QQ plots in Figure~\ref{fig:qq_gp_random}, we choose to pool together data from all three summer months due to small amounts of data.
PIT errors are displayed in Figure~\ref{fig:gp_pit}. There does not seem to be any spatial patterns in the errors,
apart from some vertical lines that appear to be artifacts of the transformation from radar reflectivity to precipitation amounts.
Overall, the absolute values of the PIT errors are generally smaller than those in Figure~\ref{fig:pit}.
This indicates that our marginal model captures the distributional tails of the data well.

Based on the presented model evaluation, we conclude that our marginal model provides a satisfactory fit to the data.

\section{Modelling the conditional intensity process}

In the main paper, when modelling the conditional intensity process, we discover that the empirical
estimates \(\hat \alpha(d; y_0)\) of \(\alpha(d)\) seem to depend on \(y_0\) for all values of
\(y_0\). We therefore propose to model \(\alpha(\cdot)\) as a function \(\alpha(d; y_0)\) that
depends on \(y_0\), and we propose the form
\begin{equation}
  \alpha(d; y_0) = \exp\left[-{\left\{d / \lambda_a(y_0)\right\}}^{\kappa_a(y_0)}\right],
\end{equation}
where \(\lambda_a(y_0), \kappa_a(y_0) > 0\) are parametric functions of \(y_0\).

\begin{figure}
  \centering
  \includegraphics[width=.99\linewidth]{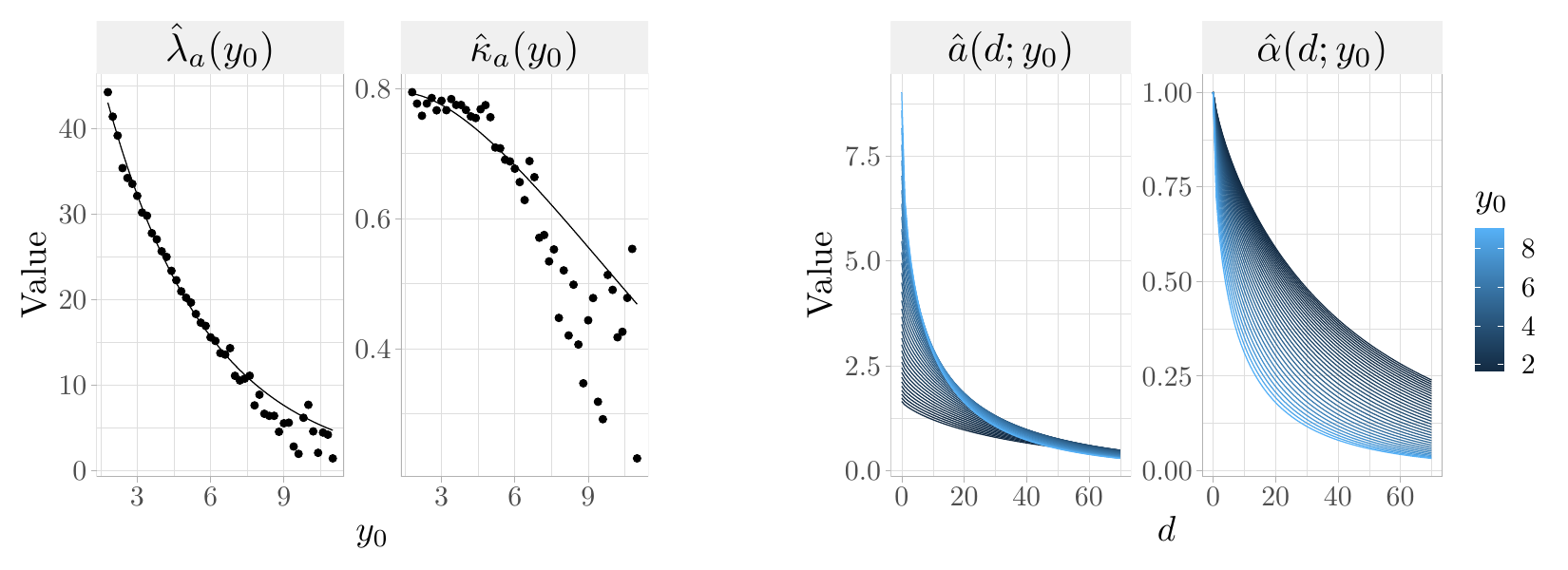}
  \caption{Leftmost plots: Points displaying least squares estimates of \(\lambda_a(y_0)\) and
    \(\kappa_a(y_0)\), computed using a sliding window over \(y_0\) with a width of \(0.2\). The
    black lines display the least squares estimates of \(\lambda_a(y_0)\) and \(\kappa_a(y_0)\)
    under the model in~\eqref{eq:a_model}, computed without the sliding window approach, i.e., using
    all possible threshold exceedances.
    Rightmost plots: The estimated functions for \(\alpha(d; y_0)\) and \(a(d; y_0)\), created using
    the least squares estimates of \(\lambda_a(y_0)\) and \(\kappa_a(y_0)\).}%
  \label{fig:a_mse_fit}
\end{figure}

Since the mean of our spatial conditional extremes model is \(a(d; y_0) = y_0 \alpha(d; y_0)\), we
can easily estimate \(\lambda_a(y_0)\) and \(\kappa_a(y_0)\) for any fixed value of \(y_0\), by
minimising the sum of squared errors between \([Y(\bm s) \mid Y(\bm s_0) = y_0]\) and
\(a(\|\bm s - \bm s_0\|; y_0)\). Thus, using a sliding window estimator over \(y_0\), with a window
size of \(0.2\), we estimate \(\lambda_a(y_0)\) and \(\kappa_a(y_0)\) for a set of different
threshold exceedances \(y_0 > \tau\). The estimators are displayed in the two leftmost plots of
Figure~\ref{fig:a_mse_fit}. We see that the estimates of \(\lambda_a(y_0)\) seem to decay
exponentially towards zero as \(y_0\) increases, while the estimates of \(\kappa_a(y_0)\) look
more like the density function of a half-Gaussian distribution. We therefore propose the
models
\begin{equation}
  \label{eq:a_model}
  \lambda_a(y_0) = \lambda_{a0} \exp(-(y_0 - \tau) / \Lambda), \quad \kappa_a(y_0) = \kappa_{a0}
  \exp(-{((y_0 - \tau) / \Lambda_\kappa)}^{\varkappa}),
\end{equation}
where the parameters \(\lambda_{a0}\), \(\Lambda\), \(\kappa_{a0}\), \(\Lambda_\kappa\) and
\(\varkappa\) all are required to be positive.

We fit the models in~\eqref{eq:a_model} to the data by once more minimising the sum of squared
errors. The resulting estimates of \(\lambda_a(y_0)\) and \(\kappa_a(y_0)\) are displayed as black
lines in the two leftmost plots of Figure~\ref{fig:a_mse_fit}, and the resulting estimates of
\(\alpha(d; y_0)\) and \(a(d; y_0)\) are displayed in the two rightmost plots of
Figure~\ref{fig:a_mse_fit}. The fitted functions seem similar to those displayed in Subplot A of
Figure~\ref{fig:empirical_conditional} in the main paper, and we therefore conclude that our chosen model for \(\alpha(d; y_0)\)
provides an adequate fit to the data.

\begin{figure}
  \centering
  \includegraphics[width=.99\linewidth]{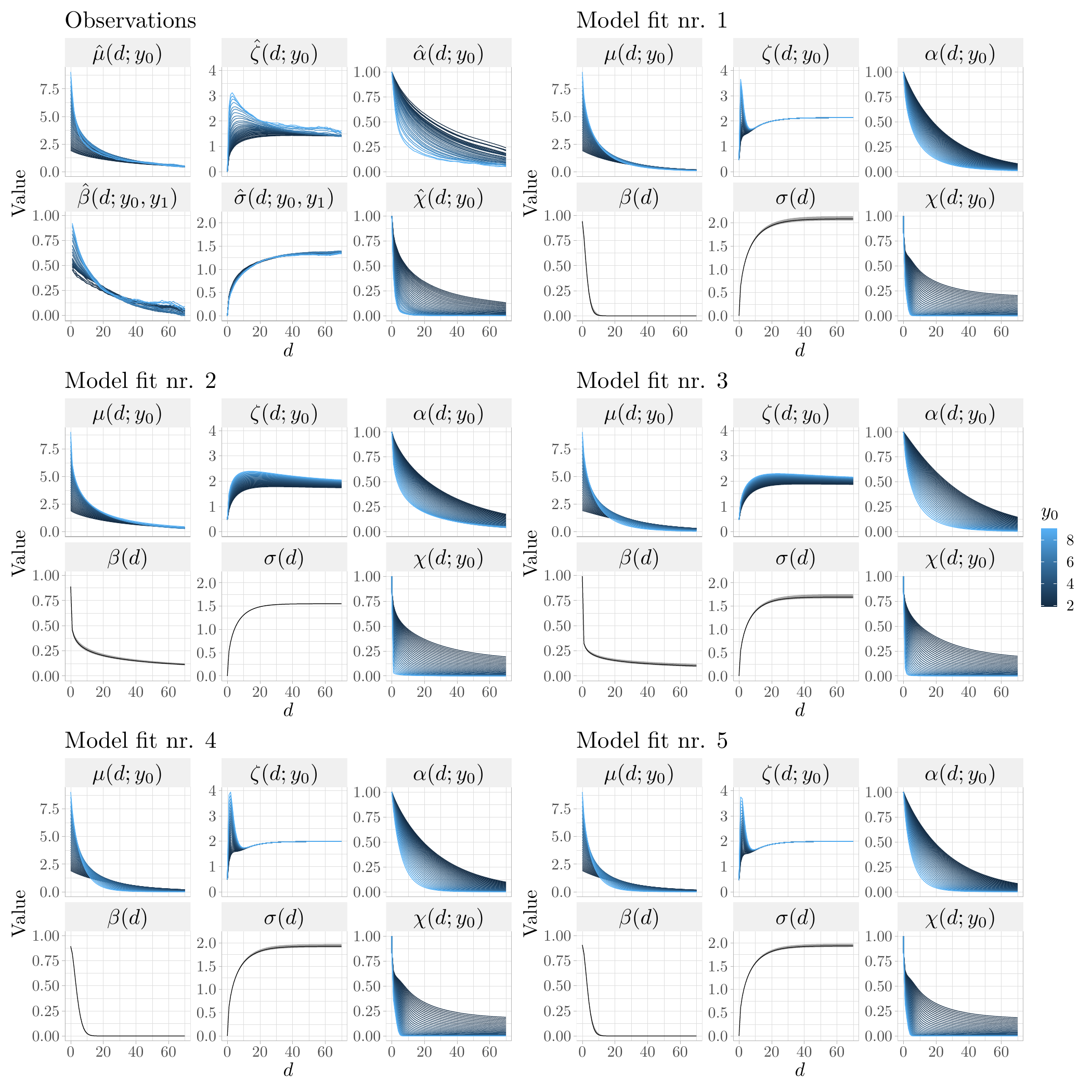}
  \caption{Estimates of \(\mu(d; y_0)\), \(\zeta(d; y_0)\), \(\alpha(d)\), \(\beta(d)\),
    \(\sigma(d)\) and \(\chi(d; y_0)\) for observed data and for simulated data using the five
    chosen conditioning sites. The estimates are computed using the same sliding window approach as
    in the main paper.}%
  \label{fig:intensity_results_appendix}
\end{figure}

Having performed inference with \texttt{R-INLA} for the conditional intensity process using each of
the five chosen conditioning sites, we evaluate model performance by simulating out-of-sample data
and computing empirical estimates of \(\mu(d; y_0)\), \(\zeta(d; y_0)\), \(\alpha(d)\),
\(\beta(d)\), \(\sigma(d)\) and \(\chi(d; y_0)\), just as in Figure~\ref{fig:empirical_conditional} in the main
paper. Figure~\ref{fig:intensity_results_appendix} displays these estimates for simulated data
based on each of the five conditioning sites. It also displays the estimates from Subplot A in
Figure~\ref{fig:empirical_conditional} of the main paper, which were computed ``globally'', by using every single location in
\(\mathcal S\) as a possible conditioning site. The estimates of \(\beta(d)\) vary slightly
between the different model fits, and this difference leads to considerable changes in the standard
deviation \(\zeta(d; y_0)\) of the model fits. These differences might be caused by the fact that,
as discussed in the main paper, our chosen model for \(\beta(d)\) is somewhat simple, in that it
does not account for the fact that the empirical estimates change as a function of \(y_0\) for
small values of \(d\). Thus, for some model fits, \(\beta(d)\) is given a value that captures the
sharp spike of \(\hat \zeta(d; y_0)\) that occurs at small \(d\) with large values of \(y_0\), while
for other fits, \(\beta(d)\) is given a value that better captures the smoother values of
\(\hat \zeta(d; y_0)\) when \(y_0\) is small. We believe that a more complex model for \(\beta(d)\),
possibly that changes as a function of \(y_0\), would be able to capture both of these
characteristics better. As shown in Section~\ref{sec:final_simulations}, the model fits with both of
the different forms of \(\beta(d)\) perform well and produce simulated data that closely capture
important properties of the observed data such as its aggregated precipitation sums and extremal
dependence structure.

\section{Evaluating the final precipitation simulations}%
\label{sec:final_simulations}

\begin{figure}
  \centering
  \includegraphics[width=.99\linewidth,page=1]{precipitation_chi.pdf}
  \includegraphics[width=.99\linewidth,page=2]{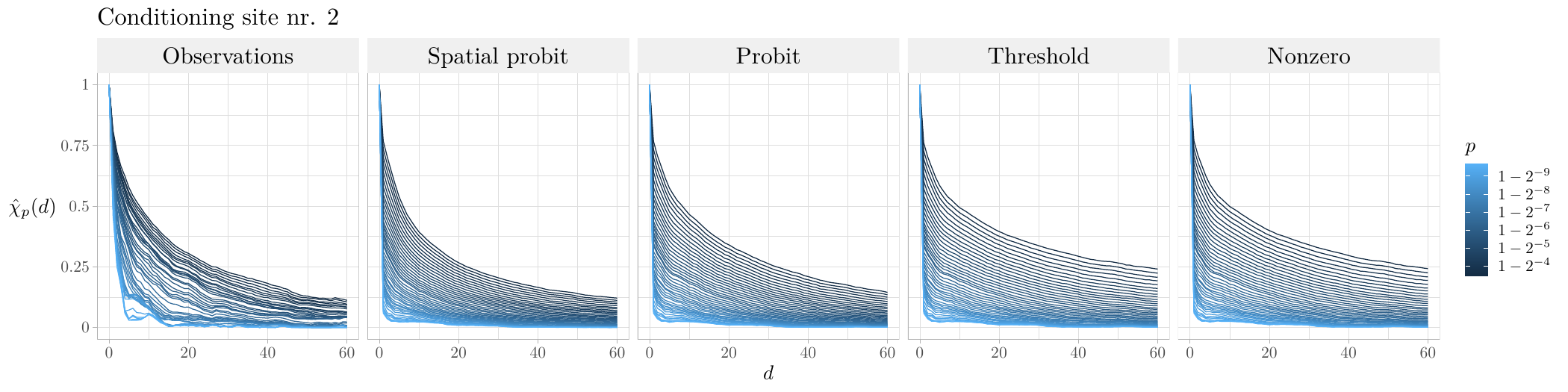}
  \includegraphics[width=.99\linewidth,page=3]{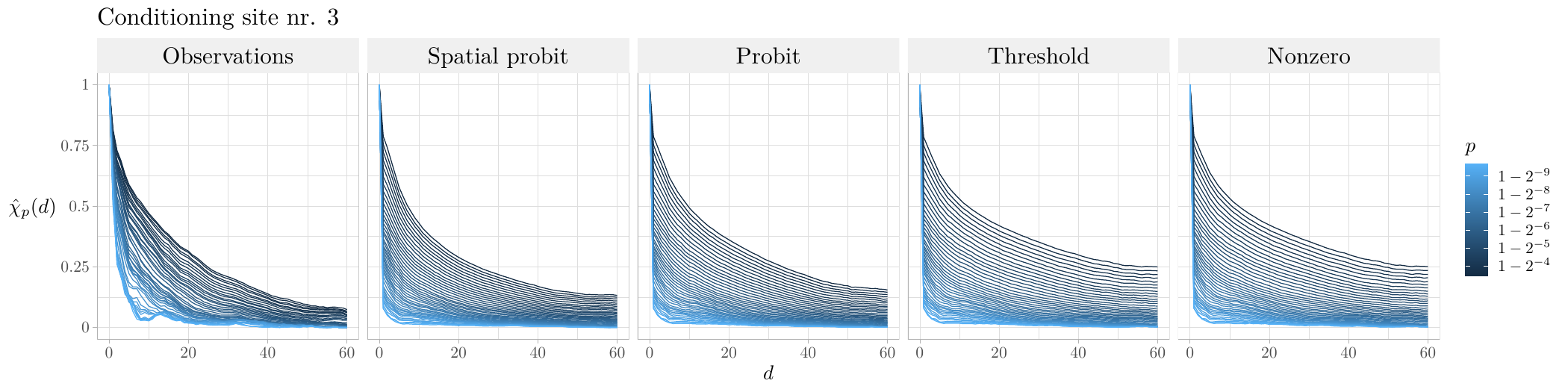}
  \includegraphics[width=.99\linewidth,page=4]{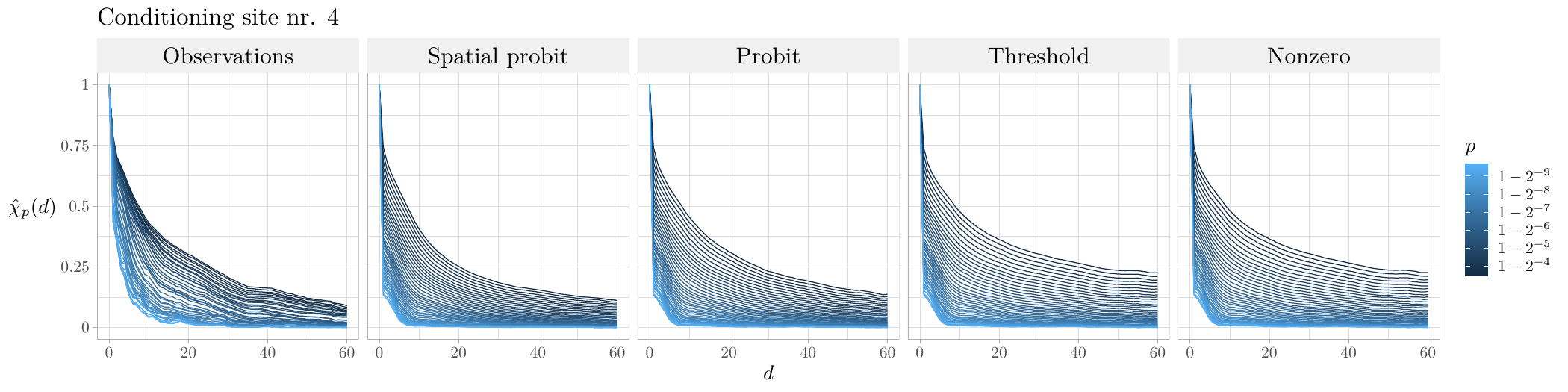}
  \includegraphics[width=.99\linewidth,page=5]{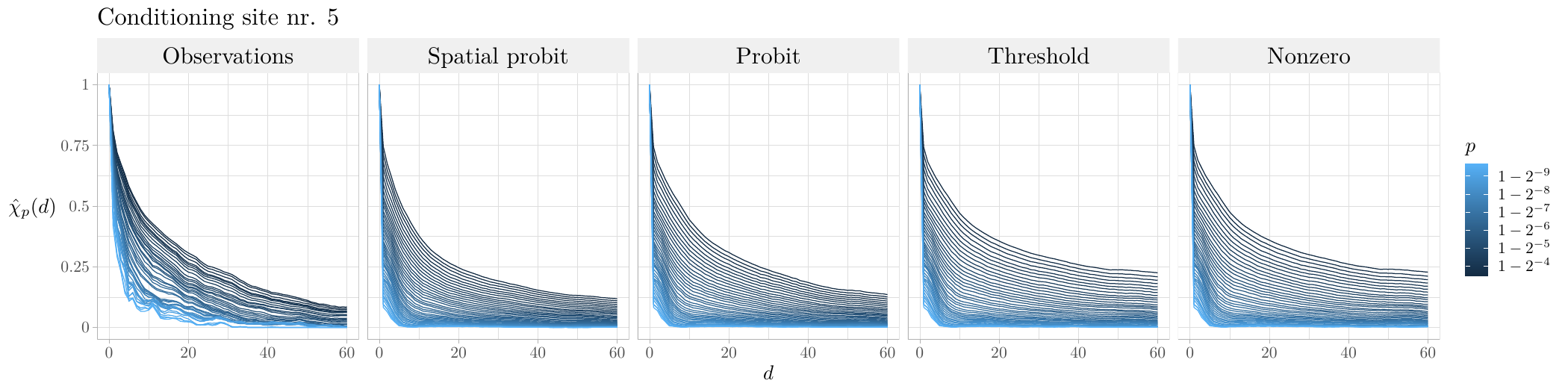}
  \caption{Estimates of \(\chi_p(d)\) for observed and simulated data, using all five conditioning sites.}%
  \label{fig:precipitation_chi_supplement}
\end{figure}

\begin{figure}
  \centering
  \includegraphics[width=.99\linewidth]{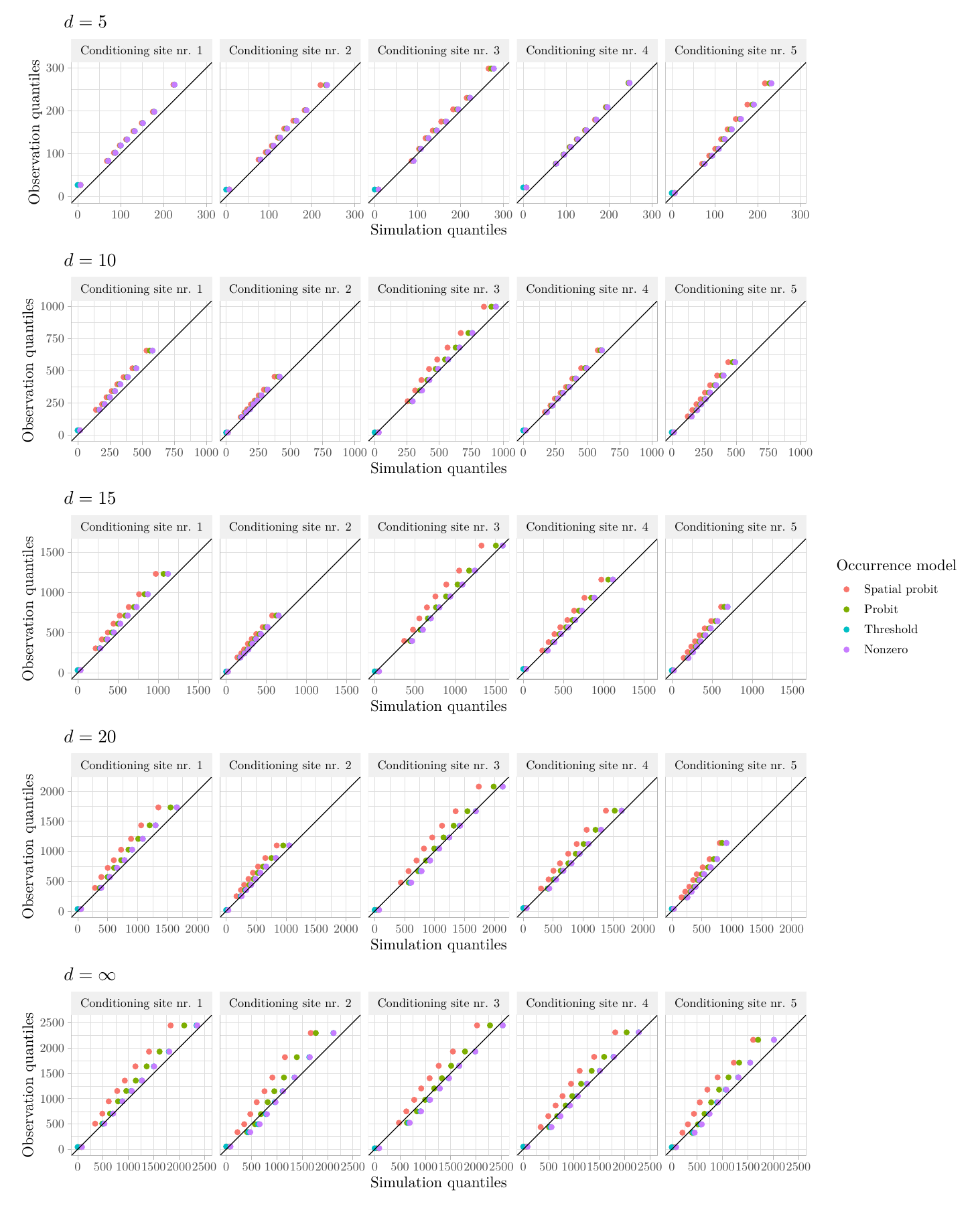}
  \caption{QQ plots for the sum of aggregated precipitation over the intersection of the
    Stordalselva catchment and a ball of radius \(d\), centred at \(\bm s_0\), for each of the five
    conditioning sites. The four different simulations are displayed using different colours.}%
  \label{fig:qq_sums_supplement}
\end{figure}

We evaluate the precipitation simulations by computing conditional exceedance probability estimates
and by creating QQ plots for the sums of aggregated precipitation over different areas inside the
Stordalselva catchment. Figure~\ref{fig:precipitation_chi_supplement} displays empirical estimates of
\(\chi_p(d)\) for both the observed and the simulated data, using each of the five conditioning
sites. The simulation-based estimates correspond well to the observation-based estimates, overall. Yet, they
tend to be smaller than those of the observed data for large \(p\) and small \(d\), and they tend to
be larger than those of the observed data for small large \(d\). This is further discussed in the
main paper.  Figure~\ref{fig:qq_sums_supplement} display QQ plots for the sum of precipitation inside a ball of
radius \(d\), centred at \(\bm s_0\), for each of the five conditioning sites, between the observed
data and simulations from the four different model fits. The QQ plots show that there is a good
correspondence between observed and simulated data, but that the simulated data tend to slightly
underestimate aggregated precipitation amounts.

\end{document}